# The Weight Is in the Reading: Hidden States in Memristors, Memlumors, and Electrochemical Transistors

**Gonzalo Rivera-Sierra,[1] Heyi Zhang,[1] Daiki Nishioka,[2] Takashi Tsuchiya,[2] Juan Bisquert[1]***

[1]Instituto de Tecnología Química (ITQ), Consejo Superior de Investigaciones Científicas-Universitat Politècnica de València, 46022, Valencia, Spain.

[2]Research Center for Materials Nanoarchitectonics (MANA), National Institute for Materials Science (NIMS), 1-1 Namiki, Tsukuba 305-0044, Ibaraki, Japan.

*Corresponding author's e-mail: jbisquer@itq.upv.es

## Abstract

Physical synapses are routinely described by a single number, the weight, that is identified with a conductance, a current, or a luminescence yield. This identification conflates two different objects: the internal state that the device stores, and the signal obtained when that state is read. Building on Chua's formulation of memristive systems, we separate the state variable, which evolves with the excitation history, from the observation map that converts it into a measured signal, and we define the weight as the outcome of a specified read protocol. Three platforms show the consequences: a memristor, in which the same voltage writes and reads; a memlumor, in which light both modifies the state and produces the photoluminescence that reports it; and an organic electrochemical transistor (OECT), in which the gate writes and the channel reads. In the transistor, two states prepared by different histories give the same channel current at a specific drain bias and become distinguishable again when the bias is changed, while the stored charge persists and the gate current continues to separate them. Published experiments provide precedents: a drain-bias-dependent cancellation of the OECT transient, and trapped charges in perovskite memlumors that are invisible to conventional time-resolved photoluminescence but revealed by a read–write–read sequence. A synaptic weight is therefore not a property of the material alone but of the material together with its readout, and reports of synaptic behavior should specify what is stored, what is measured, and how one is inferred from the other.

## 1. Introduction

How can memory disappear from a measured signal without disappearing from the material? The question is central to physical computing, which exploits the intrinsic dynamics of materials and devices to transform and retain information.[1–5] The variables that carry memory are physical: ionic distributions, filament morphologies, polarization domains, trapped charge, phase fractions, or electrochemical charge.[6,7] They are never observed directly in operation. A computational weight is always assigned from a measured signal, a current, a conductance, or an emitted light intensity, obtained under a particular terminal configuration and protocol.

This distinction is easily lost. A progressive increase of the current during a pulse train is commonly described as synaptic potentiation, and the conductance reached at the end of the train is identified with the synaptic weight.[8–10] Such a response demonstrates history dependence, but it does not establish which physical quantity evolves, whether it persists between pulses, or how it is mapped onto the signal.[11,12] Conversely, the absence of a history-dependent signal does not imply that the internal state has disappeared: the state may still be present and simply not be expressed in the quantity that is measured. Reports of hundreds to thousands of programmable conductance levels[13–16] are similarly statements about a signal under a read protocol, and the most detailed of them show that the usable levels are limited by microscopic configurations that the conductance alone does not capture.[16]

Chua and Kang formulated memristive systems as a state equation coupled to a state-dependent conductance,[17] and control theory provides the general language of state evolution, observation maps, and observability.[18,19] Reservoir computing similarly separates internal dynamics from trained readouts.[20,21] What has been missing is a consistent use of this separation in the description of physical synapses, where the weight is the quantity of interest. In this work we place the state variable, the observation map, and the read protocol at the center of the definition of a synaptic weight, and we show with three platforms that the distinction has measurable consequences.

The three platforms combine writing and reading in different ways. In a memristor the same voltage writes the state and reads it as a current.[6,17] In a memlumor, introduced by Scheblykin and co-workers, light modifies the trapped-charge state of a luminescent semiconductor and the same light produces the photoluminescence that reports it.[22,23] In an organic electrochemical transistor (OECT) the gate writes the charging state of the channel, and a separate circuit reads it through the channel current.[7,24] In all three, the expression of the stored state in the measured signal can vanish while the state persists: in the memristor when the read voltage is zero, in the memlumor under a conventional time-resolved measurement, and in the OECT at a specific drain bias with the read circuit active. Using a minimal OECT model, we show that two states prepared by different histories can give exactly the same channel current at a specific drain bias and become distinguishable again when the bias is changed, while the stored charge persists. Published experiments on OECTs[25] and on perovskite memlumors[26] provide precedents for both halves of the argument: a read condition can cancel the expression of a state, and a different read protocol can reveal a state that a conventional measurement does not see.

## 2. State, Signal, and Weight

Discussions of physical synapses often use the single word weight for quantities of different nature: a material configuration, a variable in a compact model, a measured signal, and a number used by an algorithm. Fig. 1a separates these four levels. The configuration is changed by the writing stimulus and by the kinetics of the material; the state variable is a modelling choice that must reproduce the response; the signal depends on the terminal, bias, or optical condition used to read; and the weight depends on how the measurement is processed. Figs. 1b–d show the three platforms used below.

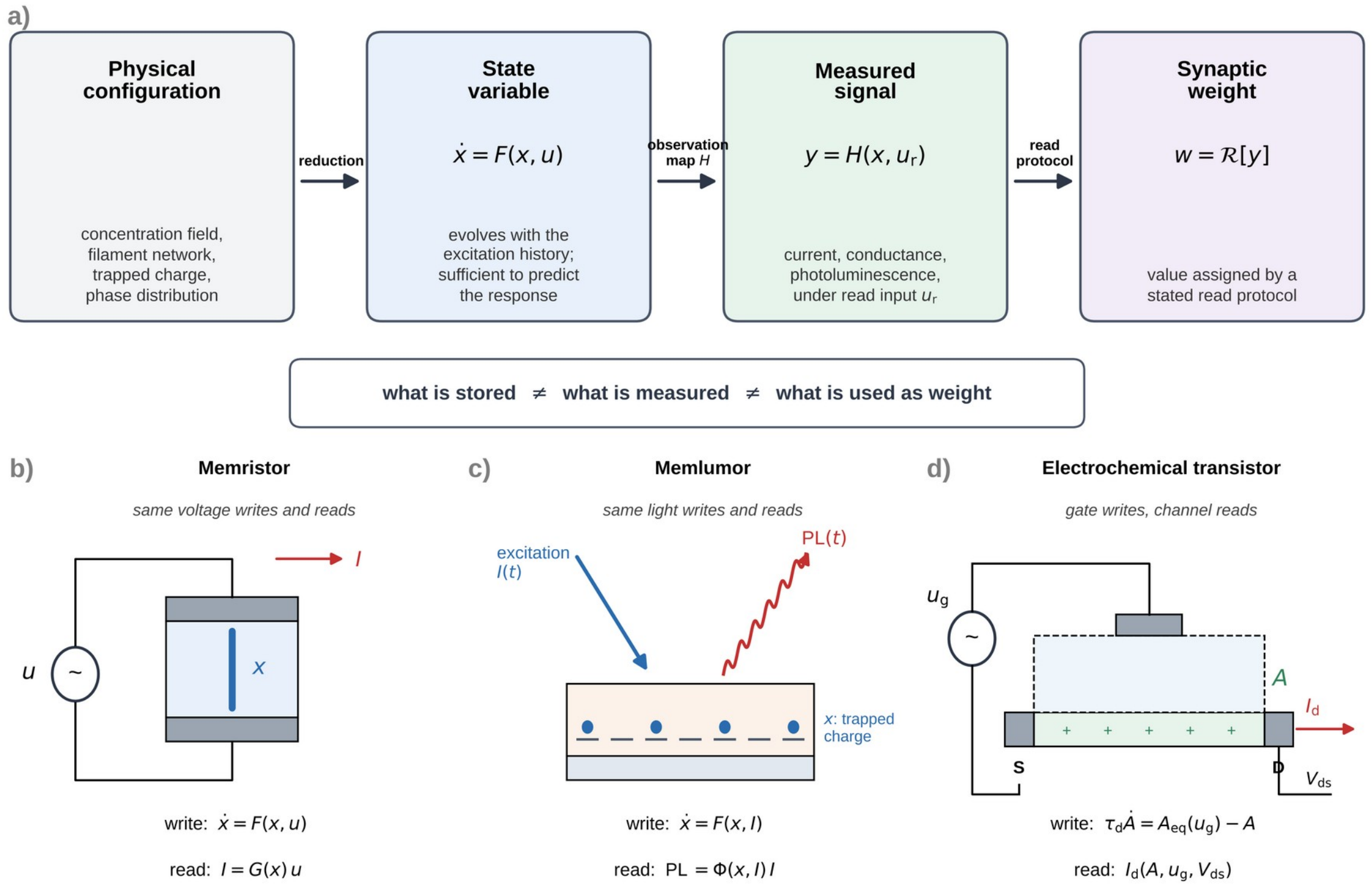

**Figure 1.** State, signal, and weight in physical synapses. (a) The physical configuration that carries memory is represented by a state variable $x$, which must be sufficient to predict the response to the specified inputs. Under a read input $u_r$, the observation map $H$ converts $x$ into a measured signal $y$, from which a weight $w$ is assigned by a read protocol $R$. (b) In a memristor the same voltage $u$ writes the state and reads it as the current $I = G(x)u$. (c) In a memlumor the excitation intensity $I(t)$ writes the trapped-charge state and reads it as the photoluminescence $PL = \Phi(x, I)I$, where $\Phi$ is the photoluminescence quantum yield. (d) In an OECT the gate voltage $u_g$ writes the channel charging state $A$, and the drain current $I_d$ at fixed drain bias $V_{ds}$ reads it.

### 2.1. State variable

The physical configuration of a memory element is generally a field or a network: a concentration profile, a filament morphology, a domain pattern, a distribution of trapped charge, or a phase distribution. A compact model replaces it by a small set of state variables $x$ obeying

$$\frac{dx}{dt} = F(x, u) \tag{1}$$

where $u$ denotes the applied stimuli, electrical or optical. The defining requirement is the state property: the present value of $x$, together with the future input, must determine the future response. If one device after two histories has the same value of a scalar variable but responds differently to the same subsequent stimulus, at least one variable is missing. Direct imaging of a filament identifies the carrier of memory, and the choice among the filament radius, the minimum neck area, or the residual gap as the state variable is then made by the descriptor that best predicts the electrical response.[27–29] In $VO_2$ a phase fraction and a defect coordinate are both required,[30] in perovskite memlumors the state contains the concentration of trapped charge and the parameters of the defects that trap it,[23] and in integrated $HfO_2$ memristors incomplete conduction channels next to the main filament control the read fluctuations, so that two states with the same mean conductance differ in the configuration that controls their noise.[16] In two-dimensional nanofluidic channels, operando interferometric imaging shows that the memristive hysteresis is produced by voltage-induced blisters of the confining walls, driven either by osmotic pressure under both polarities or by electrostatic forces under one polarity; similar current–voltage loops therefore arise from different configurations, and the effect of a given blister on the conductance depends on its location and on the other active blisters.[31] Section 3 relates the state variables of each architecture to the physical entities that generate the memory, and Table 1 summarizes representative experimental cases.

### 2.2. Observation map and read protocol

The measured signal is related to the state by an observation map,

$$y = H(x, u_r) \tag{2}$$

where $u_r$ is the input applied during the read. $H$ need not be linear, monotonic, or one-to-one, and it depends on the read input as much as on the state. Linearizing around an operating point, $\delta y = C\delta x + D\delta u$, separates two contributions. $C = \partial H/\partial x$ is the instantaneous observability of the state in the output: it measures how strongly the stored state is expressed in the signal at the given read input. $D = \partial H/\partial u$ is a direct feedthrough of the stimulus, which carries no memory. Whether the state can be reconstructed from the whole record of $y$ is the full observability problem of control theory:[18,19] a variable with vanishing $C$ may still be inferred from the time evolution of the signal if it is coupled to others that remain visible, but a single variable entering the signal affinely with $C = 0$ is not expressed at all for as long as the read condition is maintained.

A weight is then the number assigned from the complete read record $y_P$ obtained under a read protocol $P$, which fixes the terminals or optical channels, the read input, the timing, and the processing of the measurement:

$$w = R_P[y_P] \tag{3}$$

where $R_P$ is the operation that translates the read record into a synaptic weight: for example, dividing a current by the read voltage to obtain a conductance, normalizing to a reference, integrating a photoluminescence transient, or taking the ratio of two read pulses. A single current sample at a fixed read bias is the simplest record; a differential conductance, a transconductance, a time-integrated photoluminescence, or a ratio of two read pulses are different protocols, not rescalings of the same quantity. The protocol also determines which states can be told apart. Two states $x$ and $x'$ are distinguishable under $P$ when the weights obtained from them with that protocol differ by more than the resolution of the measurement, $|w(x) - w(x')| \geq \delta w_{res}$; otherwise they are equivalent for that protocol. Here $\delta w_{res}$ is the resolution of the weight, the smallest difference that can be detected with the chosen error criterion, which is set by the noise, bandwidth, and integration time of the read; when the weight is a current, we denote the corresponding current resolution by $\delta I_{res}$. In the deterministic limit the equivalence reads

$$x \equiv_P x' \Leftrightarrow y_P[x] = y_P[x'] \tag{4}$$

Changing the protocol changes this equivalence. Two physically different states can be equivalent under one protocol and distinct under another, and the number of weights that a device offers is a property of the device together with its protocol. Reading may also perturb the state, because the read input enters Eq. (1); a nondestructive read is an approximation whose validity must be shown. The three architectures of Fig. 1 differ precisely in how the read input is related to the write input.

## 3. Three Architectures

### 3.1. Memristor: one voltage writes and reads

In a two-terminal memristor[17] the state variable $x$ evolves toward a voltage-dependent equilibrium $x_{eq}(u)$ with a characteristic time $\tau_x(u)$,[12,32]

$$\frac{dx}{dt} = \frac{x_{eq}(u) - x}{\tau_x(u)} \tag{5}$$

and the current is

$$I = G(x)u \tag{6}$$

where $G(x)$ is the state-dependent conductance. Depending on the device, $x$ may describe filament connectivity, an interfacial vacancy distribution, a domain fraction, or an ion-concentration profile (Table 1).[27,33–35] The applied voltage plays two coupled roles (Fig. 1b): it drives Eq. (5) and converts the conductance into current. The instantaneous observability of the state, $\partial I/\partial x = u\, dG/dx$, vanishes only when the probe is removed, $u = 0$, and the probe that restores it also drives the state. Every read is therefore a weak write. The read voltage must be small or brief enough not to change $x$ appreciably, but large enough that the signal separates as many states as possible above the noise, so that it is a compromise between read disturbance and resolution. The model functions and parameters are given in Section S1 of the Supporting Information. The conductance is a convenient weight because it is an invertible function of $x$, but its meaning as a state rests on independent evidence for the configuration that it summarizes. Simulations of programming and reading with this model are shown in Section S2.

### 3.2. Memlumor: one light beam writes and reads

Scheblykin and co-workers introduced the memlumor, a luminophore with memory, as the optical counterpart of the memristor.[22] The excitation intensity $I(t)$ plays the role of the voltage, and the photoluminescence plays the role of the current,

$$PL(t)=\Phi(x,I)I(t), \frac{dx}{dt}=F(x,I) \quad (7)$$

where the photoluminescence quantum yield $\Phi$ depends on the excitation history through the state $x$. In $CsPbBr_3$ films and microcrystals, $\Phi$ changes on time scales from nanoseconds to minutes, with switching energies down to a few femtojoules.[22] Within a Shockley–Read–Hall description, the state contains the concentration of trapped charge carriers, which photodope the semiconductor and increase the radiative recombination of subsequently generated carriers, together with slower variables describing the photoinduced dynamics of the trap states themselves.[23] Eq. (7) has the same structure as Eqs. (1) and (6), and $\Phi(x,I)$ is the optical observation map. As in the memristor, the light that reads the state also writes it: every excitation pulse creates carriers that can be trapped. The quantum yield is often identified with the synaptic weight,[22] but what is measured is a photoluminescence signal under a specific excitation sequence, and different sequences express the same trapped-charge state differently. Memlumors have been used to classify binary sequences of light pulses from the time-integrated photoluminescence,[36] which is a weight defined by a specific protocol in the sense of Eq. (3).

### 3.3. Electrochemical transistor: the gate writes and the channel reads

In the accumulation-mode OECT, the gate voltage $u_g$ controls the ionic–electronic charging state of the channel. With the charging variable $A$, defined as the number of injected electronic charges per unit channel length so that $qLA$ is the stored charge,[37,38]

$$\tau_d \frac{dA}{dt}=A_{eq}(u_g)-A \quad (8)$$

where $A_{eq}(u_g)$ is the gate-defined equilibrium and $\tau_d$ the ionic charging time.[39] The ionic charging current supplied through the gate is

$$I_g=qL\frac{dA}{dt}=\frac{qL}{\tau_d}\left[A_{eq}(u_g)-A\right] \quad (9)$$

and the compensating electronic charge enters the channel through both contacts, a fraction $f$ through the drain and $1-f$ through the source.[24] With the transport current $-\theta q L A/\tau_e$, where $\tau_e$ is the electronic transport time and $\theta = \pm 1$ its sign relative to the charging flux, the drain and source currents are[37,40]

$$I_d = -\theta \frac{qL}{\tau_e} A - f I_g, I_s = -\theta \frac{qL}{\tau_e} A + (1-f) I_g \tag{10}$$

which satisfy current conservation, $I_s = I_d + I_g$ (Section S3). Writing and reading are now separated (Fig. 1d): the gate controls the state through $\tau_d$, whereas its expression in the channel current also depends on the drain circuit through $\tau_e$. In the linear regime $\tau_e \approx L^2/(\mu |V_{ds}|)$, with $\mu$ the electronic mobility,[24] so that the drain bias tunes the observation map. Varying $\tau_e$ at fixed $A_{eq}$ and $\tau_d$ is a controlled theoretical experiment; in a real device the drain bias can also modify the local gate–channel potential and hence the charging itself. In the simulations $A$ is normalized by the reference charging state $A_{ref} = A_{eq}(-0.5V)$ and the currents by $I_0 = q L A_{ref}/\tau_d$; for a PEDOT:PSS channel of typical dimensions $I_0$ is of the order of nanoamperes (Section S1).

The state–readout distinction also applies to electric-double-layer transistors (EDLTs). OECTs couple ion uptake in the channel to volumetric doping,[41] whereas EDLTs in the electrostatic gating regime use interfacial ionic redistribution to modulate channel carrier density. This electrostatic coupling has been demonstrated in solid-electrolyte/hydrogenated-diamond EDLTs, where channel carrier modulation was quantified by Hall measurements.[42] The ionic configuration contributes to the internal state, while its expression in the measured current depends on the channel transfer characteristic and drain bias. In graphene EDLTs, transient terminal currents can additionally include contributions associated with double-layer charging.[43] Thus, distinguishing ionic-state dynamics from electrical readout extends the framework to EDLT-based artificial synapses and reservoirs.

Table 1 collects representative devices of the three classes in which the physical configuration has been identified, together with the state variables used to describe them and the signal from which the weight is assigned.

**Table 1.** Physical configuration, state variable, and readout in representative devices. Extended compilations are given in Tables S2 and S3.

| Device | Physical configuration and evidence | State variable(s) | Signal and weight |
|---|---|---|---|
| Ag/$SiO_2$ and Ag/a-Si electrochemical metallization memristor[27] | Metallic filament imaged by in situ and ex situ TEM; size set by compliance current | Gap length, minimum cross-section, or connected volume | Read current; conductance |
| Pt/$TiO_2$/Pt valence-change memristor[28] | Magnéli-phase filament identified by HRTEM | Filament connectivity or cross-section | Read current; conductance |
| $HfO_2$/$Al_2O_3$ memristor integrated on CMOS[16] | Main channel plus incomplete channels resolved by conductive AFM; trapped charge causes telegraph noise | Conductance plus at least one hidden coordinate controlling fluctuations | Current in a 0–0.2 V read sweep; 2048 levels after denoising |
| Co/Au/$BaTiO_3$/LSMO ferroelectric tunnel-junction memristor[44] | Voltage-controlled polarization domains | Up-domain area fraction | Tunnelling current; conductance |
| $VO_2$ memristor (volatile and nonvolatile)[30] | Phase and disorder evolution by in situ X-ray nanoimaging | Metallic fraction and defect coordinate | Read current; conductance |

| Device | Physical configuration and evidence | State variable(s) | Signal and weight |
|---|---|---|---|
| Two-dimensional hBN nanofluidic channel memristor[31] | Voltage-induced blistering of the confining walls imaged operando by interferometry; osmotic (bidirectional) or electrostatic (unidirectional) origin | Blister height and area; adhesion state | Ionic current; conductance |
| $CsPbBr_3$ perovskite memlumor[22,23,26] | Trapped photogenerated charge and photoinduced trap dynamics; read–write–read photoluminescence and recombination modelling | Trapped-charge concentration; trap and recombination parameters | Photoluminescence under a read pulse; quantum yield or time-integrated emission |
| PEDOT:PSS OECT (transistor)[24,45] | Cation uptake and dedoping; established electrochemistry | Channel charge | Drain or source current at fixed drain bias; volatile weight |
| $PdH_x$/Nafion/$WO_3$ protonic synaptic transistor[14] | Proton insertion quantified by charge accounting, XPS and XAS | Hydrogen stoichiometry | Drain current at 0.1 V; conductance over seven decades |
| PEDOT:PSS ionic floating-gate memory (transistor)[46] | Polymer redox state isolated by a volatile Ag selector | Redox charge and selector connectivity | Read current below 10 nA; conductance |
| Crystalline–amorphous OECT (transistor)[15] | Selective ion doping of morphologically distinct regions | Ion populations of two regions | Drain current under a standardized read; retained conductance |

# 4. Hidden and Recovered States

In all three architectures the observability of the stored state in the measured signal can vanish while the state persists. What differs is the read condition at which this happens, and whether the state can be recovered without being disturbed. We first state the conditions for the three devices, then discuss experimental cases, and finally show the consequence for two prepared states.

## 4.1. Conditions for hiding a state

In the memristor, the instantaneous observability of the state is

$$C_M = \frac{\partial I}{\partial x} = u \frac{dG}{dx} \tag{11}$$

which vanishes at $u = 0$. Between programming pulses the state relaxes while the current is zero, and its evolution leaves no trace in the signal; a small read voltage makes it visible and separable again (Section S2, Fig. S1). The condition is trivial, but it is the same situation analyzed below: the state persists while its observability vanishes. Here the price of recovery is a read that also drives the state.

In the memlumor, the instantaneous observability of the state in the photoluminescence follows from Eq. (7),

$$C_L = \frac{\partial PL}{\partial x} = I \frac{\partial \Phi}{\partial x} \tag{12}$$

which, as in the memristor, vanishes without excitation, $I = 0$: trapped charge evolves between pulses without being expressed in the emission, and each excitation pulse that makes it visible also writes it. In addition, the memlumor shows that visibility can depend on the whole excitation sequence and not only on an operating point. Under a periodic train of identical pulses the accumulated decay does not separate the trapped charge from the other photoexcitations,[26] although $C_L$ is finite during each pulse, whereas a sequence of isolated read pulses around a write burst does separate it (Section 4.2).

In the OECT, the observability can vanish at a finite read bias, with the read circuit active. Inserting Eq. (9) into Eq. (10), the current at terminal $k$ is

$$I_k = -\frac{qL}{\tau_d}\left[\left(\theta\frac{\tau_d}{\tau_e} - s_k f_k\right)A + s_k f_k A_{eq}(u_g)\right] \tag{13}$$

with $s_d f_d = f$ for the drain and $s_s f_s = -(1-f)$ for the source. The first term is the state contribution and the second a feedthrough of the gate voltage. The instantaneous observability of the stored charge at terminal $k$ is

$$C_k = \frac{\partial I_k}{\partial A} = -\frac{qL}{\tau_d}\left(\theta\frac{\tau_d}{\tau_e} - s_k f_k\right) \tag{14}$$

which vanishes at a finite bias when

$$\theta\frac{\tau_d}{\tau_e} = s_k f_k \tag{15}$$

Because $\tau_d/\tau_e > 0$, the condition is met at the drain for $\theta = +1$, at $\tau_d/\tau_e = f$, and at the source for $\theta = -1$, at $\tau_d/\tau_e = 1-f$: the transport current and the share of the charging current cancel at the one terminal at which they have opposite signs. At the compensating drain the current reduces to

$$I_d = -\frac{qLf}{\tau_d}A_{eq}(u_g) \tag{16}$$

and is independent of $A$. The channel continues to charge and relax according to Eq. (8), and the gate current of Eq. (9) continues to reflect it, but the history is not represented in $I_d$ for as long as the read condition is maintained. Under a pulse train the drain current then follows the gate waveform without pulse-to-pulse accumulation, and the forward and reverse branches of the transfer curve coincide, whereas on either side of compensation the accumulation and the hysteresis reappear with opposite signs (Section S2, Fig. S2; complete voltage-sweep and pulse series for both polarities and for depletion-mode devices are given in Sections S6\u2013S9). A vanishing hysteresis alone does not prove that the state is invisible; here this is established by the coefficient of Eq. (14).

## 4.2. Experimental precedents

Experiments on the three classes of devices show that whether a stored state is observed depends on how it is read. In integrated $HfO_2$ memristors, incomplete conduction channels next to the main filament are not observable in a single conductance read, which returns only the mean conductance, but they appear in the statistics of repeated reads as random telegraph fluctuations and in conductive atomic force microscopy; two states with the same mean conductance therefore differ in a hidden variable that controls their noise, and suppressing it by denoising pulses increased the number of usable levels.[16] In the OECT, the compensation has a direct experimental counterpart. Friedlein et al. measured the source current of PEDOT:PSS OECTs after a small gate-voltage step at different drain voltages (Fig. 2a).[25] At −30 mV the current jumped and then decayed toward its new value, at −80 mV it became an essentially instantaneous step, and at −130 mV a smaller jump was followed by a gradual rise, while the gate current displayed the same ionic transient in all cases. The authors attributed the fast response to the partition of the gate current between the contacts. The reduced model reproduces the three shapes with the single parameter $|\theta|\tau_d/\tau_e$ (Fig. 2b): below the compensation value the current overshoots, at compensation it is a pure step, and above it the current rises with the charging time. Because $\tau_e$ decreases as $|V_{ds}|$ increases, a larger drain voltage crosses the compensation point in the order observed. The measured terminal is the one at which the two contributions have opposite signs, as required by Eq. (15). The comparison concerns the shape of the responses; a quantitative description would use the device parameters of ref. 25.

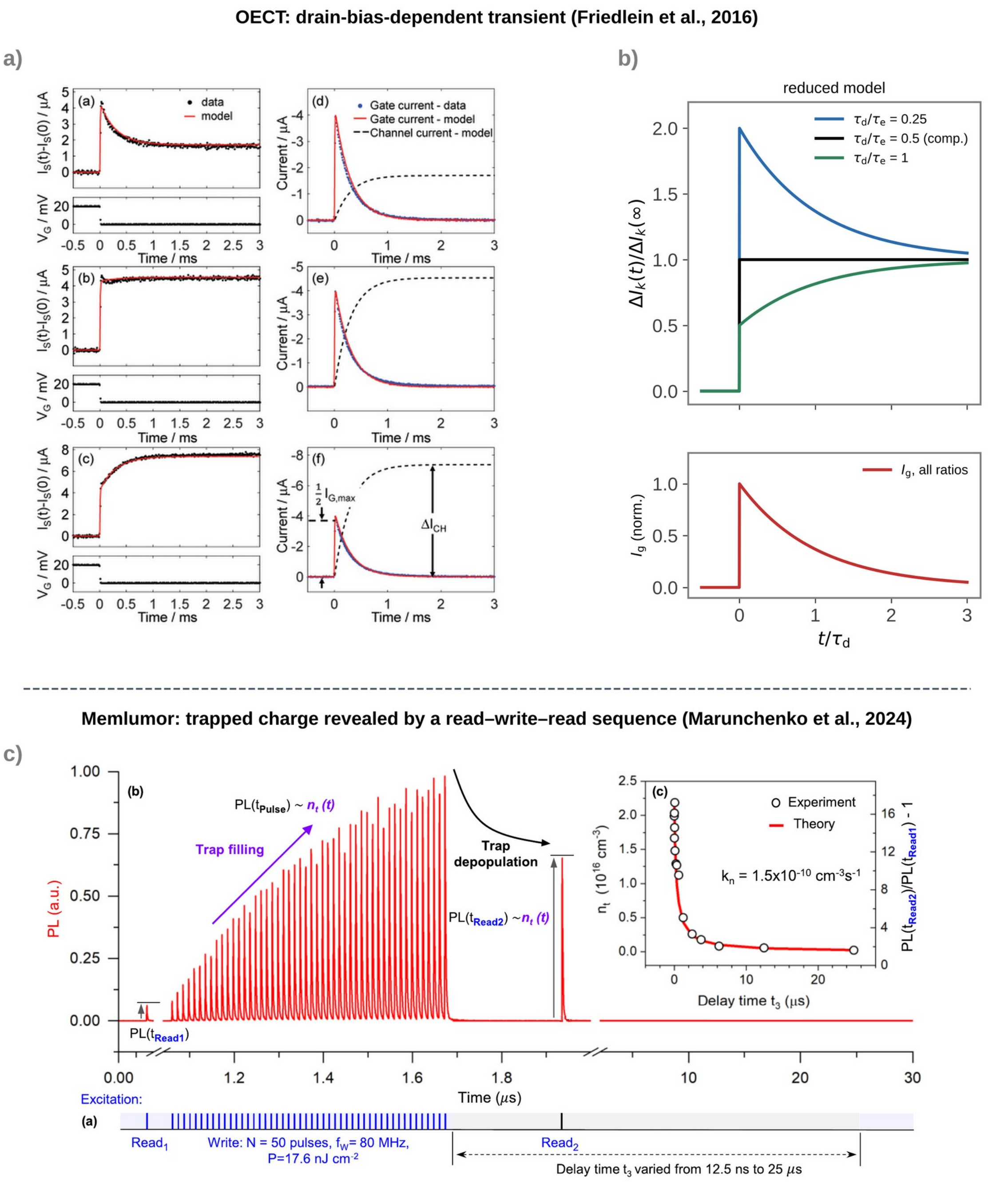

**Figure 2.** Experimental precedents for protocol-dependent visibility of a stored state. (a) Change of the source current of a PEDOT:PSS OECT after a gate-voltage step from 20 mV to 0 at drain voltages of −30, −80, and −130 mV (inner panels a–c), and the corresponding gate and modelled channel currents (inner panels d–f). Reproduced from ref. 25; copyright 2016 Wiley-VCH (permission to be confirmed). (b) Reduced model, Eq. (13), at the compensating terminal for $|\theta|\tau_d/\tau_e=0.25$, 0.5, and 1 with $f_k=0.5$, normalized to the final change, and the gate current, Eq. (9), identical for all ratios. (c) Photoluminescence of a $CsPbBr_3$ microplate under a read–write–read sequence: a single read pulse (Read$_1$), a burst of 50 write pulses at 80 MHz, and a second read pulse (Read$_2$)

after a variable delay (inner panel a). The emission grows during the burst as charges are trapped, and the second read is stronger than the first (inner panel b); the ratio of the two reads gives the concentration of trapped charge and its decay with the delay (inner panel c). Reproduced from ref. 26, published under a CC-BY 4.0 license.

The memlumor provides the optical counterpart (Fig. 2c). Marunchenko et al. showed that conventional time-resolved photoluminescence, which accumulates the decays produced by a periodic train of identical pulses, cannot directly detect photoexcited species that live much longer than the photoluminescence decay, such as trapped charge in $CsPbBr_3$.[26] These species are present in the material and modify the response, but the averaged decay does not express them in a separable form. A read–write–read protocol makes them visible: a single weak pulse reads the initial state, a burst of pulses writes it by filling traps, and a second single pulse reads it again after a controlled delay. The ratio of the two read signals measures the concentration of trapped charge, about $2\times10^{16}$ $cm^{-3}$ after the burst, and its decay with the delay gives the rate of trap depopulation.[26] In the language of Section 2, the trapped-charge state is equivalent to the ground state under the conventional protocol and distinct from it under the read–write–read protocol. The mechanism differs from the OECT compensation, but the lesson is the same: whether a stored state is seen depends on how it is read.

### 4.3. Two prepared states read under identical conditions

The operational consequence of compensation appears when two different states must be compared, which is the basic task of reading a stored weight. Two charging states are written into the OECT by different gate histories, twelve and three pulses ending at the same time $t_p$ and at the same gate voltage (Fig. 3a), and are then read under identical conditions, changing only the drain bias. They leave the channel in different charging states (Fig. 3b), with $\Delta A(t_p)=0.096\,A_{ref}$. Because the gate voltage during the read is common to both, the feedthrough term of Eq. (13) cancels in the difference, and the contrast between the two histories is

$$\Delta I_d(t)=-\frac{qL}{\tau_d}\left(\theta\frac{\tau_d}{\tau_e}-f\right)\Delta A(t),\Delta I_g(t)=-\frac{qL}{\tau_d}\Delta A(t) \qquad (17)$$

To decide whether this contrast is detectable, it is compared with the current resolution $\delta I_{res}$ defined in Section 2.2, the smallest current difference that the read can resolve with the chosen error criterion. In Fig. 3 we take $\delta I_{res}=0.01\,I_0$, and we examine below how the conclusions change with the resolution.

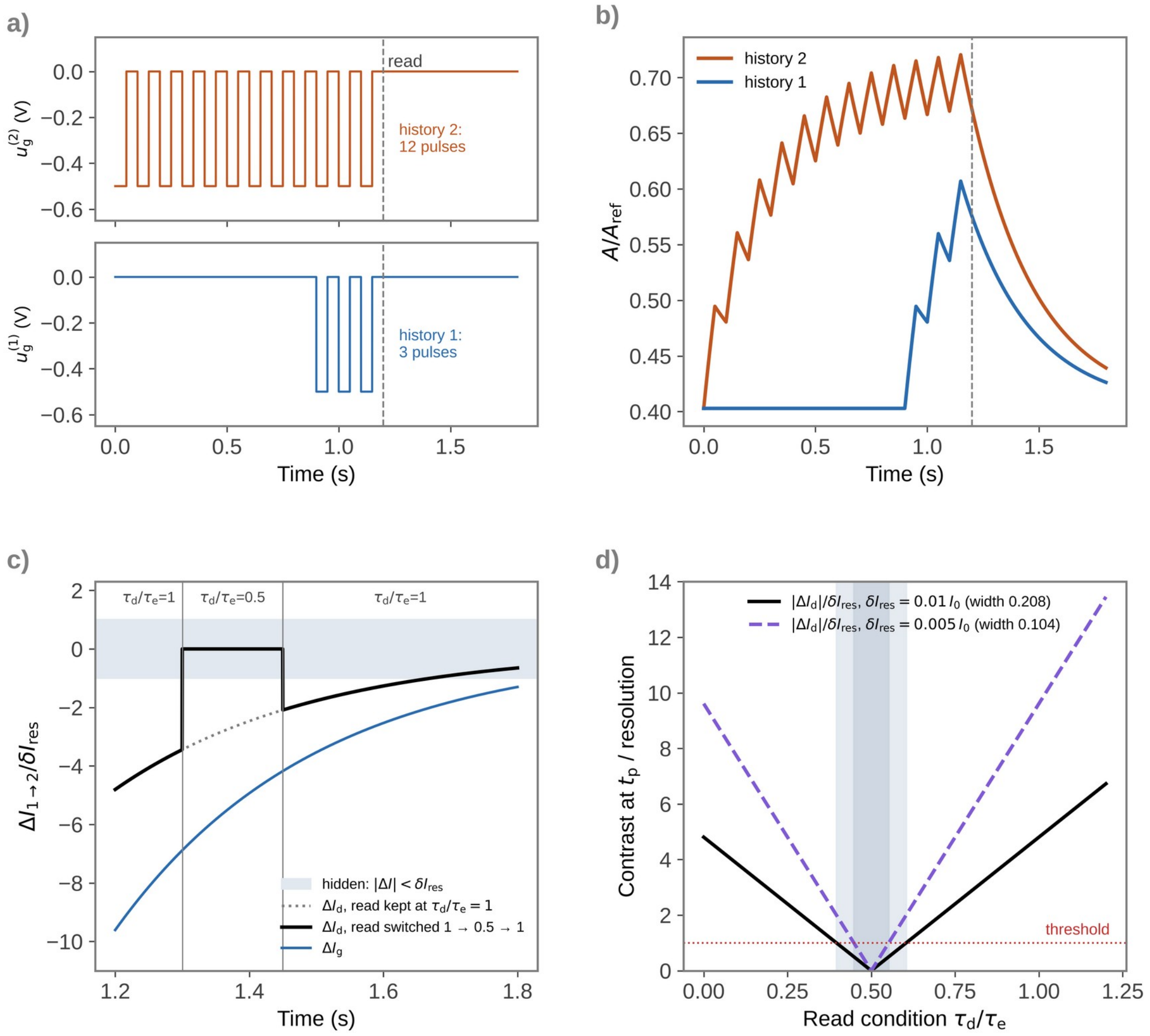

**Figure 3.** Two charging states whose current contrast is hidden and recovered by the read condition alone ($\theta=+1$, $f=0.5$, $\tau_d=0.3$ s; currents in units of $I_0=qLA_{ref}/\tau_d$). (a) Gate histories 2 (top, twelve −0.5 V pulses) and 1 (bottom, three pulses), both ending at $t_p=1.2$ s, after which the gate is held at 0 V and the device is read. (b) Corresponding charging trajectories. (c) Contrast between the two histories divided by the resolution $\delta I_{res}=0.01\,I_0$: drain current when the read condition is switched from $\tau_d/\tau_e=1$ to the compensation value 0.5 and back at 1.45 s (solid), drain current read at 1 throughout (dotted), and gate current (blue); the shaded band marks contrasts below the resolution. (d) Contrast at $t_p$ as a function of the read condition for $\delta I_{res}=0.01\,I_0$ and $0.005\,I_0$; shaded regions mark the practical invisibility of Eq. (18).

At $\tau_d/\tau_e=1$ the drain-current contrast is $0.048\,I_0$, about five times the resolution $\delta I_{res}=0.01\,I_0$ assumed in Fig. 3, and the gate contrast is $0.096\,I_0$. Switching the drain bias to the compensation condition removes the drain-current contrast (Fig. 3c): the two histories become equivalent under this read, although $\Delta A$ remains finite and the gate current keeps distinguishing them. Restoring the original bias brings the contrast back to $0.021\,I_0$, the value of a read kept at the original condition throughout. What is recovered is the readout contrast; the state difference itself relaxes continuously, and recovery must be judged against the relaxation during the hidden interval. With a finite resolution, the histories cannot be distinguished within a region around the compensation point (Fig. 3d),

$$\left|\theta\frac{\tau_d}{\tau_e}-f\right|<\frac{\tau_d\delta I_{res}}{qL|\Delta A|} \tag{18}$$

whose full width in $\tau_d/\tau_e$ is 0.21 for $\delta I_{res}=0.01I_0$ and 0.10 for $0.005I_0$. A coarser resolution widens the region and eventually hides the histories at every read condition.

The same experiment has counterparts in the other two architectures. In the memristor, two states with different $x$ give the same zero current at $u=0$ and become distinguishable at a finite read voltage, at the price of perturbing both. In the memlumor, a trapped-charge state and the ground state are not separated by the averaged decay under periodic excitation and are separated by the ratio of two read pulses (Fig. 2c). The OECT is distinctive because the contrast vanishes at a finite read bias, with the read circuit active and without perturbing the state, and is recovered by changing only that bias.

Exact cancellation is a property of the affine single-variable model. In a real device the partition factor and the ionic time may depend on bias and state, and several charging populations with different times cannot all be compensated by one read condition. Section S10 shows that a slow population carrying up to 20% of the charge, or a partition factor that varies by up to 50% across the charging range, leaves a residual contrast below the resolution assumed in Fig. 3 (Fig. S11). The experimental statement is therefore that the contrast is not resolvable under a calibrated protocol. A direct test follows from Fig. 3: two histories ending at the same gate condition are read near and away from compensation while the gate current and one channel current are recorded, and a reference–compensation–reference sequence is compared with an unswitched read at matched elapsed times.

## 5. Consequences for Physical Synapses

Memristors, memlumors, and electrochemical transistors are very different devices. They store memory in filaments or vacancy distributions, in trapped photogenerated charge, and in the electrochemical charge of a polymer channel, and they are written and read with voltages, with light, and with separate gate and channel circuits. They are nevertheless often treated as black boxes whose value lies in reproducing temporal patterns of biological synapses, such as potentiation, paired-pulse facilitation, or fading memory.[8–10,22] The analysis presented here shows that these patterns have a common origin and a common structure. In every case a physical–chemical process changes an internal configuration, which is described by state variables, and in every case the synaptic response is the projection of those variables onto a signal under chosen inputs and read conditions. The questions that determine the usefulness of a device as a synapse are therefore the same for all of them: which internal state is written, how that state can be accessed, and how the inputs and the read condition are controlled. Placing the three devices in one framework separates what belongs to the material, the kinetics of its internal state, from what belongs to the measurement, the observation map and the protocol, and makes devices of very different physics comparable.

The three architectures make explicit a distinction that is general. In the memristor and the memlumor, the input that reads the state also writes it, so that the state can be hidden only by removing the probe, and every read perturbs it. In the OECT, writing and reading are separated, and the read condition becomes a design variable that can enhance, suppress, or restore the expression of a stored state without rewriting it. The memlumor shows in addition that the protocol, not only the operating point, decides visibility: the same material and the same trapped charge are invisible or quantitatively measured depending on the sequence of excitation pulses.

The drain circuit can also be used to control the state dynamics themselves. Shibata et al. demonstrated physical masking in a $LiCoO_2$-based redox ion-gating reservoir by applying stepped triangular drain voltages, which increased the diversity of measured reservoir features and improved computational performance.[47] A later study by Nishioka et al. reported an increase in total information-processing capacity from 11 to 20, mainly through enhanced nonlinear capacity, with little change in linear memory capacity under the conditions examined.[48] In the latter study, the conductance modulation was interpreted in terms of a changing effective gate voltage that drives additional ionic dynamics.[48] Physical masking thus illustrates a regime in which the read-circuit bias can influence both state evolution and its measured expression, extending readout engineering beyond changes in the observation map under fixed state dynamics.

Three consequences follow. First, a synaptic weight is a property of the material together with its readout. Reported potentiation, conductance levels, or quantum-yield changes are statements about a signal under a protocol, and their physical interpretation requires the observation map. Second, the absence of a signal is not the absence of a state: a device can retain information that its chosen readout does not express, and a different terminal, bias, or optical sequence may recover it. Third, the read protocol can be engineered. Studies that reduce read noise[16] or decouple the read path from drift[49] have enlarged the readable state space of memristors without changing the material, and the OECT compensation suggests that a three-terminal device can modulate the visibility of its stored contrast; because the feedthrough of Eq. (16) remains, exploiting this in arrays would require baseline subtraction or differential sensing. Hiding a state also differs from preserving it: its visibility can be suppressed while it continues to relax.

For experimental reports, these considerations translate into the minimum information summarized in Table 2. Multilevel reports specify the read protocol and the statistical criterion to different extents,[13–15,46,50–52] and the number of levels that can actually be distinguished, and for how long, depends on the noise, the decision rule, and the relaxation of the state. These quantitative questions build on the distinctions made here and will be treated separately.

**Table 2.** Information required to interpret a reported synaptic state.

| Level | What to report | Why it matters |
|---|---|---|
| Physical configuration | Evidence for the memory carrier (structural, chemical, spectroscopic, charge accounting) or inference from the response | Separates mechanism from model |
| State variables | Variables used to describe the device and the conditions under which they predict its response | Reveals hidden variables that affect noise or retention |
| Read signal | Terminal or optical channel and measured quantity (current, conductance, transconductance, photoluminescence); read input; delay, duration, bandwidth, pulse sequence | Defines what is measured and its sensitivity to the state |
| Weight extraction | Normalization, ratio, integration, differential or calibrated procedure | Defines the weight under the stated protocol |
| Levels | Number of programmed targets; repeated readings of each level; criterion used to distinguish adjacent levels | Distinguishes programmed from readable levels |
| Read disturbance | Evidence that reading does not change the stored state over the required number of reads | Validates a nondestructive readout |
| Time dependence | Coding scheme; number of distinguishable levels at stated hold times and temperature | Distinguishes retention of the material state from retention of usable levels |

## 6. Conclusions

A material memory becomes a synaptic weight only through a specified read. Building on Chua's formulation of memristive systems, we separated the physical configuration, the state variable, the measured signal, and the weight, and defined the equivalence of states under a read protocol. We then used three very different artificial synapses, a memristor, a memlumor, and an electrochemical transistor, to show that their synaptic behavior follows from one common structure: an internal state written by a physical–chemical process, and a signal that expresses that state only to the extent allowed by the inputs and the read condition. In all three the observability of the state can vanish while the state persists, in the memristor when the read voltage is removed, in the memlumor under a conventional time-resolved measurement, and in the transistor at a finite-bias compensation condition. In the transistor, two states written by different histories become indistinguishable in the channel current and distinguishable again when only the read is changed, while the stored charge persists and the gate current continues to separate them. Published experiments on memristors, OECTs, and perovskite memlumors show that the visibility of a stored state is set by the read condition and the protocol, whatever the physical nature of the state.

What matters for a physical synapse is therefore not only what it stores but how the stored state is accessed. Describing a device requires stating what is stored, what is measured, and how one is inferred from the other. This separation turns the readout into a design variable, provides a common basis for comparing electrical and optical synapses, and calls for experiments that combine the electrical or optical readout with operando probes of the physical configuration, as recently achieved for nanofluidic memristors.[31]

## Supporting Information

Supporting Information is available from the publisher: reduced model functions in dimensionless form, parameters, protocols, and numerical methods (Section S1, Table S1); programming, common read, and null readout in the reduced models (Section S2); terminal currents and current conservation in the OECT (Section S3); compilations of identified internal states in memristors and ionic synaptic transistors (Sections S4 and S5, Tables S2 and S3); voltage-sweep projections and accumulation- and depletion-mode OECT pulse series (Sections S6–S9); robustness of the compensation (Section S10); and the scripts used to compute all figures (Section S11).

## Acknowledgments

The work was funded by the European Research Council (ERC) via Horizon Europe Advanced Grant, grant agreement nº 101097688 ("PeroSpiker"). The work was in part supported by Japan Science and Technology Agency (JST) as part of Adopting Sustainable Partnerships for Innovative Research Ecosystem (ASPIRE), Grant Number JPMJAP2530. Funding for open access charge: CRUE-Universitat Politècnica de València.

## AI Use Statement

During the preparation of this work the authors used ChatGPT (OpenAI) and Claude (Anthropic) to assist with literature organization, critical reading of drafts, language editing, and the writing of simulation and plotting scripts. All models, derivations, results, and interpretations were checked by the authors, who take full responsibility for the content of the publication.

## Associated content

Data Availability Statement

The data presented here can be accessed at https://doi.org/10.5281/zenodo.XXXX (Zenodo) under the license CC BY 4.0 (Creative Commons Attribution 4.0 International).

# Supporting Information

# The Weight Is in the Reading: Hidden States in Memristors, Memlumors, and Electrochemical Transistors

**Gonzalo Rivera-Sierra,[1] Heyi Zhang,[1] Daiki Nishioka,[2] Takashi Tsuchiya,[2] Juan Bisquert[1]***

[1]Instituto de Tecnología Química (ITQ), Consejo Superior de Investigaciones Científicas-Universitat Politècnica de València, 46022, Valencia, Spain.

[2]Research Center for Materials Nanoarchitectonics (MANA), National Institute for Materials Science (NIMS), 1-1 Namiki, Tsukuba 305-0044, Ibaraki, Japan.

*Corresponding author's e-mail: jbisquer@itq.upv.es

# S1. Reduced models, numerical conventions, and parameter definitions

The main-text equations define the state evolution and electrical observation maps; the explicit functions, parameters, initial conditions, and numerical methods are given below so that every figure can be reproduced. Time constants are reported in milliseconds in Table S1 and in seconds on the time axes of the figures. All OECT simulations, in the main text and here, use the dimensionless charging coordinate $A/A_{ref}$ and currents in units of $I_0$, as defined in Section S1.1.

**Table S1. Key numerical parameters and voltage protocols used in the main-text and supplementary simulations.**

| Quantity | Fig. S1 | Fig. S2a,b | Supplementary part |
|---|---|---|---|
| Electrical protocol | Positive memristor pulses; negative OECT gate pulses | OECT forward–reverse gate sweep | Memristor/OECT sweeps and positive/negative OECT gate pulses |
| OECT polarity factor, $\theta$ | -1 | +1 | -1 and +1 |
| Ionic charging time, $\tau_d$ | 300 ms | 300 ms | 300 ms |
| Electronic transport time, τe | 50 ms | varied | 50, 600, and 900 ms |
| Time-scale ratio, $r=\tau_d/\tau_e$ | 6 | 0 ¿$r\leq$ 1.5 (plotted) | 6, 0.50, and 0.33 |
| Relaxation weighting factor, f | 0.50 | 0.50 | 0.50 |
| OECT gate-pulse amplitude | -0.50 V | — | -0.50 and +0.50 V |
| Gate-voltage sweep range | — | -0.50 → +0.50 → -0.50 V | same range |
| Sweep-rate parameter, $V_r$ | — | 0.05, 0.20, and 0.50 $Vs^{-1}$ | 0.20 $Vs^{-1}$ in the transfer-series panels |
| Displayed state coordinate | x and A/A_ref | A/A_ref | A/A_ref or M/M_ref; currents in units of I0 |

### S1.1. Model functions, units, and initial conditions

**Memristor.** The equilibrium state of Eq. (5) of the main text and the conductance of Eq. (6) are

$$x_{eq}(u)=\left[1+\exp\left(-\frac{u-u_0}{V_m}\right)\right]^{-1},G(x)=g_L+(g_H-g_L)x \tag{S1}$$

with $u_0=0.1$ V and $V_m=0.05$ V. The relaxation time is voltage independent in all simulations, $\tau_x = 0.3$ s, and the limiting conductances are $g_L = 10$ µS and $g_H = 4$ mS. The initial condition is $x(0) = x_{eq}(0)$. Programming pulses are 0.5 V, 50 ms on and 50 ms off; in Fig. S1 they are followed by 10 ms read pulses at 50 mV every 100 ms. Because $x_{eq}(50\,mV) \neq x_{eq}(0)$, each read perturbs the state slightly.

**OECT, accumulation mode.** The gate-defined equilibrium of Eq. (8) of the main text is written in dimensionless form as

$$\frac{A_{eq}(u_g)}{A_{ref}} = \left(\frac{u_v - u_g}{u_v + 0.5\,V}\right)^{3/2} \tag{S2}$$

with $u_v = 0.6$ V and $A_{ref} = A_{eq}(-0.5\,V)$; the $3/2$ power is that of a parabolic density of states filled up to the level set by the gate. $A$ is the number of injected electronic charges per unit channel length, $L$ the channel length, and $q\,L\,A$ the stored charge. Currents are expressed in units of $I_0 = q\,L\,A_{ref}/\tau_d$, so that Eqs. (9)–(10) of the main text contain only $A/A_{ref}$, $\theta$, $f$, and $r = \tau_d/\tau_e$. In all simulations $\tau_d = 0.3$ s and $f = 0.5$. The initial condition of pulse simulations is $A(0) = A_{eq}(0)$; sweeps start at equilibrium at −0.5 V, proceed to +0.5 V at constant rate $V_r$, and return.

**Illustrative physical scale.** For a PEDOT:PSS channel with volumetric capacitance $C^{¿} \approx 40$ F cm$^{-3}$, width 100 µm, thickness 100 nm, and length 10 µm, a gate excursion of 0.5 V stores a charge $Q = C^{¿}\,W\,d\,L\,\Delta V \approx 2 \times 10^{-9}$ C. With $\tau_d = 0.3$ s this gives $I_0 \approx Q/\tau_d \approx 7$ nA, so that the resolutions $\delta I_{res} = 0.01\,I_0$ and $0.005\,I_0$ used in Fig. 3 of the main text correspond to tens of picoamperes. These values set the order of magnitude only and are not a fit to a particular device.

**OECT, depletion mode.** The equilibrium depletion coordinate of Eq. (S9) is, in the same dimensionless form,

$$\frac{M_{eq}(u_g)}{M_{ref}} = \left|\frac{u_g - E_V}{0.5\,V - E_V}\right|^{3/2} \tag{S3}$$

with $E_V = -1$ V and $M_{ref} = M_{eq}(0.5\,V)$; the reference population in Eq. (S8) is $P_0 = 1.5\,M_{ref}$, larger than the maximum depletion reached in the protocols, and the depletion currents are in units of $q\,L\,M_{ref}/\tau_d$.

**Electronic transport time.** In the linear regime the electronic transit time is[1]

$$\tau_e \approx \frac{L^2}{\mu |V_{ds}|} \tag{S4}$$

where $\mu$ is the electronic mobility; Eq. (S4) holds for drain voltages small compared with the gate overdrive. In the simulations $\tau_e$ is set directly through $\tau_d/\tau_e$ while $A_{eq}$ and $\tau_d$ are kept fixed. This corresponds to a controlled theoretical experiment; in a real device the drain bias can also modify the local gate–channel potential, so that charging and readout may be partly coupled.

### S1.2. Numerical methods and figure-specific protocols

All model figures, Fig. 2b and Fig. 3 of the main text and Figs. S1–S11, were computed in Python with an exact exponential update for piecewise-constant inputs, , with $\Delta t=0.25$ ms for pulse protocols and $\Delta t=\Delta u/V_r$ on a 3000-point voltage grid for sweeps; the scripts (make_figs4.py, make_figs6.py, and make_si_figs.py) are provided with the Supporting Information. The signed hysteresis area of Eq. (S5) is obtained by trapezoidal integration of the forward and reverse branches on a common voltage grid.

In Fig. 3 of the main text, history 1 consists of three −0.5 V pulses starting at 0.9 s and history 2 of twelve pulses starting at 0 s, both ending at $t_p=1.2$ s; the gate is then held at 0 V. In Fig. 3c the read condition is $\tau_d/\tau_e=1$ for 1.2–1.3 s, 0.5 for 1.3–1.45 s, and 1 afterwards, with instantaneous switching. Fig. 3d evaluates Eq. (17) of the main text for the state difference at $t_p$, $\Delta A(t_p)=0.096\,A_{ref}$, which is equivalent to reading replicated trajectories at a single condition each. Fig. 2b is analytical: after a gate step at $t=0$, $A(t)/A(\infty)=1-e^{-t/\tau_d}$. Fig. S2c,d uses the gate-pulse protocol of Fig. S1d (−0.5 V, 50 ms on and off, 12 pulses) with $\theta=+1$.

## S2. Programming, common read, and null readout in the reduced models

Fig. S1 compares the memristor and the OECT during programming with comparable pulse timing and during a subsequent read at fixed conditions. In the memristor, successive pulses (Fig. S1a) increase $x$ (Fig. S1c) and the current peaks (Fig. S1b). Between pulses the terminal voltage vanishes and the current is zero, so the inter-pulse relaxation of $x$ leaves no signature in the current, because the sensitivity $\partial I / \partial x$ vanishes at $u=0$. During the read phase, small pulses recover the decaying conductance (Fig. S1b, inset). The read voltage also drives Eq. (5) of the main text: $x_{eq}(50\,mV)$ differs from $x_{eq}(0)$, so each read slightly perturbs the state, which is the read-disturbance trade-off of the two-terminal architecture.

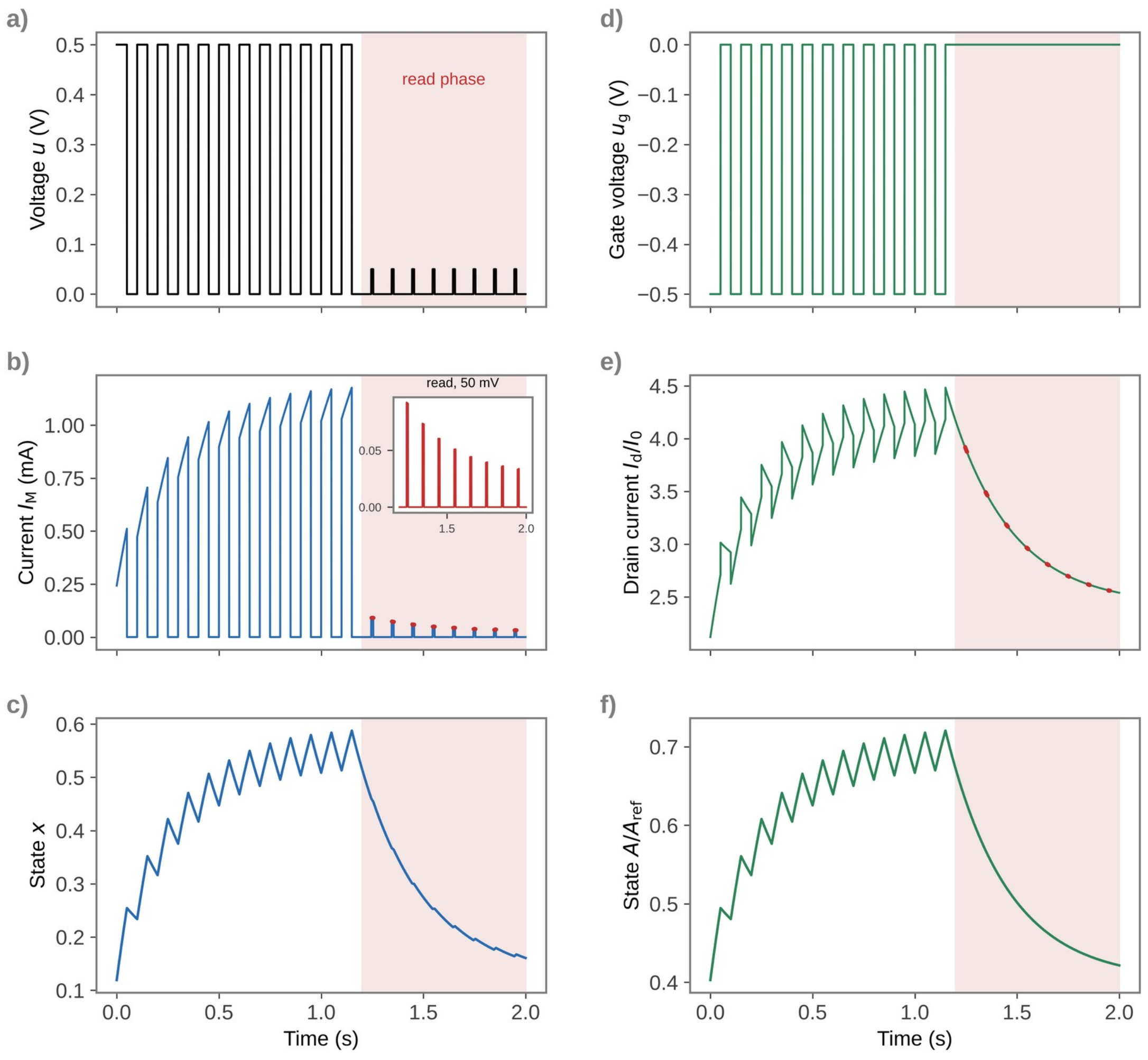

**Figure S1.** Programming followed by a common read. (a) Voltage applied to the memristor: twelve 0.5 V pulses, then 10 ms read pulses at 50 mV every 100 ms (shaded read phase). (b) Memristor current; read samples in red, enlarged in the inset. (c) State $x$.

(d) OECT gate voltage: twelve −0.5 V pulses, then the gate held at 0 V. (e) Drain current in units of $I_0$ for $\theta=-1$, $\tau_d=0.3$ s, $\tau_e=0.05$ s, with read samples at the times of the memristor reads. (f) Charging variable $A/A_{ref}$.

In the OECT, the gate pulses (Fig. S1d) charge the channel (Fig. S1f). Because the drain bias remains applied, the drain current is finite between pulses (Fig. S1e), and after programming it follows the relaxation of $A$ continuously. The peak currents during writing contain the feedthrough term of Eq. (13) of the main text, whereas the read samples at fixed gate voltage provide the weights. The two devices show similar potentiation, although the states, the read paths, and the reasons for the inter-pulse currents differ.

Fig. S2 illustrates the compensation condition of Eq. (15) of the main text. At compensation the forward and reverse transfer branches coincide, whereas away from it the delayed state opens a loop (Fig. S2a). Fig. S2b shows the signed hysteresis area

$$S_H=\int_{-0.5V}^{0.5V}\left[I_{d,FS}(u_g)-I_{d,RS}(u_g)\right]du_g \quad \text{(S5)}$$

where $I_{d,FS}$ and $I_{d,RS}$ are the drain currents on the forward (increasing $u_g$) and reverse sweeps, for three sweep rates. The rate changes the magnitude of the hysteresis, but all curves cross zero at the same ratio, and the area changes sign on either side of it. In the time domain (Fig. S2c), the drain current at compensation follows the gate waveform without pulse-to-pulse accumulation, whereas on either side the accumulation of $A$ reappears with opposite signs. Fig. S2d shows that the charging variable keeps accumulating and the gate current displays a transient at every pulse edge, while the drain current carries no trace of the history.

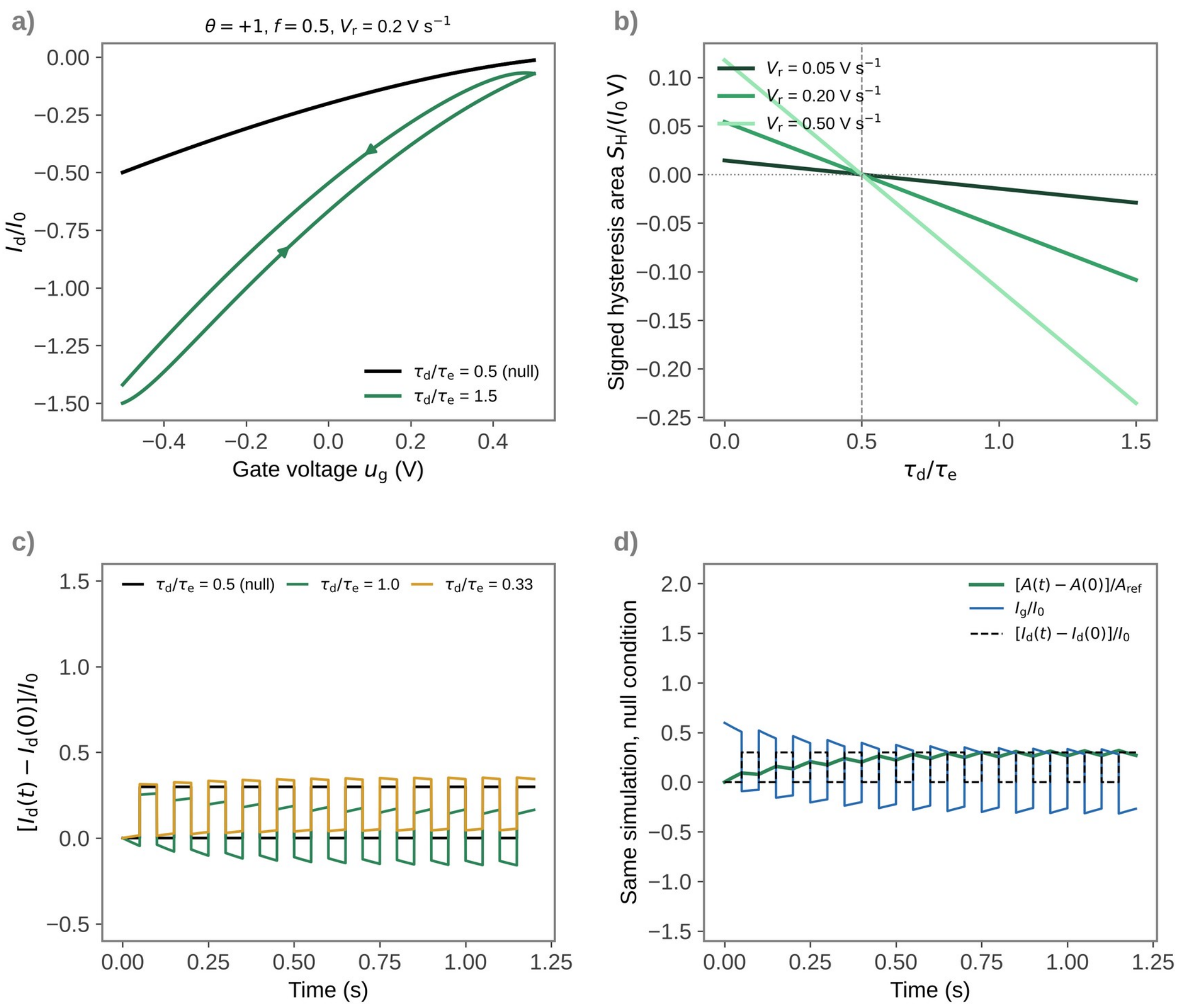

**Figure S2.** Null readout at the drain of the accumulation-mode OECT ($\theta=+1$, $f=0.5$, $\tau_d=0.3$ s; currents in units of $I_0$). (a) Transfer curves for a forward–reverse gate sweep from −0.5 V at $V_r=0.2$ V s$^{-1}$; at $\tau_d/\tau_e=0.5$ both branches coincide, whereas a loop opens at 1.5. (b) Signed hysteresis area, Eq. (S5), as a function of $\tau_d/\tau_e$ for three sweep rates. (c) Drain-current response to the pulse train of Fig. S1d at, and on either side of, the compensation condition. (d) One simulation at the compensation condition: charging variable, gate current, and drain current.

## S3. Terminal currents and current conservation in the OECT

The reduced OECT model contains three currents: the ionic charging current through the gate circuit, $I_g=qL\,dA/dt$, and the electronic currents at the two channel contacts. The electronic charge that compensates the injected ions enters through both contacts, a fraction $f$ through the drain and $1-f$ through the source,[1] while the transport current $-\theta qLA/\tau_e$ flows through the channel from one contact to the other. Defining $I_d$ as the current leaving the channel at the drain, $I_s$ as the current entering it at the source, and $I_g$ as the charging current leaving the channel toward the gate circuit, Eq. (10) of the main text satisfies

$$I_s=I_d+I_g \tag{S6}$$

Both terminals carry the same transport contribution and opposite-signed shares of the charging current. Substituting Eq. (9) of the main text gives

$$I_d = -\frac{qL}{\tau_d}\left[(\theta r - f)A + f A_{eq}\right], I_s = -\frac{qL}{\tau_d}\left[(\theta r + 1 - f)A - (1 - f)A_{eq}\right] \tag{S7}$$

with $r = \tau_d / \tau_e > 0$. The coefficient of $A$ vanishes at the drain only for $\theta = +1$, at $r = f$, and at the source only for $\theta = -1$, at $r = 1 - f$. For a given sign of the transport current relative to the charging flux, the compensation can therefore occur at one terminal only, the one at which the two contributions have opposite signs, while at the other terminal they add. Friedlein et al. measured the source current and attributed its fast response to the partition of the gate current between the contacts;[2] their measured terminal is the compensating one. The main-text analysis with the drain and $\theta = +1$ is algebraically identical, with $f$ in place of $1 - f$; for an equal partition, $f = 1/2$, the two compensation conditions coincide numerically. The partition factor depends on geometry and bias, so the compensating ratio is in general different for the two terminals of a real device.

## S4. Physical identity of the internal state in memristors

The reduced memristor model treats $x$ as an effective internal coordinate. Table S2 compiles representative experimental cases in which the physical carrier of memory has been identified by direct structural, chemical, or spatial evidence, distinguishing that evidence from mechanisms inferred mainly from electrical behavior. In several classes the microscopic carrier is convincingly identified, but the observed microstate is usually multidimensional: microscopy reveals a spatial field or morphology, whereas a compact model uses a scalar. Direct observation of a metallic or oxygen-deficient filament identifies the memory carrier, but it does not decide whether the best scalar is the filament radius, the minimum neck area, the residual gap, the reduced volume, or a percolation variable; the appropriate reduced coordinate is the geometrical feature that dominates $G(x)$. Ferroelectric tunnel junctions are an unusually clean case, in which the area fraction of one polarization orientation acts directly as a scalar state variable while the tunnelling resistance is its electrical readout.[3] Filamentary and ferroelectric devices thus represent two limits that illustrate why the conductance is a readout of the internal state rather than the state itself.

**Table S2.** Representative experimentally supported physical realizations of the memristor state variable $x$.

| State type | System | Candidate reduced coordinate | Evidence identifying the state | Ref. |
|---|---|---|---|---|
| Metal-atom distribution, filament connectivity | Ag/a-Si/W and Ag/$SiO_2$ cells | Minimum filament cross-section; gap length; connected Ag volume | In situ and ex situ TEM with compositional analysis during programming; filament size set by compliance current | 4 |
| Reduced crystalline filament | Pt/$TiO_2$/Pt | Connectivity or cross-section of the Magnéli-phase filament | HRTEM identification of the filament phase; formation and disruption correlated with transport | 5 |

| State type | System | Candidate reduced coordinate | Evidence identifying the state | Ref. |
|---|---|---|---|---|
| Oxygen-deficient core–shell filament | Pt/$HfO_2$/Pt | Continuity and size of the Hf6O core | Atomic-resolution imaging of complete and ruptured filaments; metallic versus semiconducting transport | [6] |
| Main channel plus incomplete channels | $HfO_2$/$Al_2O_3$ memristor integrated on CMOS | Mean conductance plus a coordinate for incomplete channels and trapped charge | Conductive AFM of conduction channels before and after denoising; telegraph noise linked to incomplete channels | [7] |
| Interfacial vacancy distribution | Pt/$TiO_2$−x/Pt | Vacancy density near the interface; barrier-modulation coordinate | Engineered vacancy profiles, switching localization, polarity control, barrier-type transport | [8] |
| Ferroelectric domain fraction | Co/Au/$BaTiO_3$/ LSMO ferroelectric tunnel junction | Area fraction of up/down polarization domains | Voltage-controlled domain configurations; resistance represented by parallel domain fractions | [3] |
| Charged domain-wall position | Ultrathin BiFeO3 film | Wall position; fraction of switched layers | In situ biasing STEM, EELS, and electron holography correlated with resistance states | [9] |
| Amorphous/crystalline phase configuration | Ge2Sb2Te5 phase-change cell | Crystalline fraction; amorphous-plug volume | In situ TEM under DC and nanosecond pulses resolves reversible structural transitions | [10] |
| Metallic/insulating phase and defects | $VO_2$ element | Metallic-phase fraction plus a defect coordinate | In situ X-ray nanoimaging of electrically driven phase and disorder evolution | [11] |
| Ion-concentration profile | Nanocapillary nanofluidic memristor | Enrichment/depletion amplitude | Electrical hysteresis correlated with fluorescent tracking of ion-concentration polarization | [12] |

## S5. Internal state, readout, and weight in ionic synaptic transistors

For an ionic synaptic transistor the minimal description is $dx/dt=F(x,V_G,t)$, $I_D=H(x,V_{DS},V_G)$, and $w=R(I_D;read\,protocol)$. Under an off-gate read at a small fixed drain voltage $V_r$, $I_r=G(x)V_r$ and the assigned weight is commonly $w=I_r/V_r=G(x)$: an operational, not a microscopic, identification. Across the strongest examples in Table S3, $x$ is an ionic composition, a redox or doping fraction, or a spatial ion distribution in the channel. These quantities are related by electroneutrality and redox chemistry but are not always interchangeable. A single scalar is most defensible when insertion is spatially homogeneous and the Faradaic efficiency is known; during fast switching, under strong drain bias, or with phase separation or several reservoirs, $x$ is more properly a concentration field or a vector of occupancies. The literature generally understands the mechanism correctly but compresses the three levels into the expressions "synaptic weight" or "postsynaptic current". Three situations make the distinction important: during a gate pulse, $I_D(t)$ contains simultaneous state evolution and instantaneous transistor response; a weight based on peak postsynaptic current depends on the drain and gate voltages, sampling delay, and pulse duration; and the independent drain circuit permits a genuinely low-disturbance read only if the gate is returned to a defined condition and the drain bias does not redistribute ions.

**Table S3.** Physical state, readable quantity, and assigned weight in representative ionic synaptic transistors.

| **Physical state** $x$ | **Device** | **Write and state dynamics** | **Readable quantity and weight** | **Evidence; ref.** |
|---|---|---|---|---|
| Channel cation charge; PEDOT oxidation fraction | Aqueous PEDOT:PSS OECT | Gate-driven cation uptake dedopes PEDOT; ions relax out after the pulse (volatile) | $I_D(t)$ at fixed $V_{DS}$; weight usually peak or residual current | Ionic–electronic transient model; established OECT chemistry [1,13,14] |
| Stored ionic/Faradaic charge; channel redox fraction | PEDOT:PSS ENODe | Charge transferred between reservoir and channel, retained at open circuit | $I_r$ at fixed $V_{DS}$; $w=G$ | Electrochemical charge control; nonvolatile multilevel $G$ [15] |
| Polymer redox charge plus selector connectivity | Ionic floating-gate PEDOT:PSS with Ag selector | Selector enables charge transfer, then ruptures to isolate the redox state | Low read current; weight assigned as $G_{SD}$ | Redox mechanism plus independently characterized selector [16] |
| Hydrogen stoichiometry and proton profile | $PdH_x$/Nafion/$WO_3$ | Metered H+ and e− intercalate to form Hx$WO_3$ | $I_D$ at 0.1 V; $w=G(x_H)$ | Charge accounting, XPS/XAS, calculations; $G$ versus $x_H$ [17] |
| Li occupancy in $Li_yCoO_2$ | $LiCoO_2$ solid-electrolyte transistor | Gate extracts or inserts Li+, changing channel phase and transport | Small-bias channel current; $w=G$ | Known intercalation chemistry; state mostly inferred from charge [18] |
| Cu concentration and distribution | $CuO_x$/$HfO_x$/$WO_3$ transistor | Progressive Cu migration and insertion under gate pulses | Channel $I_D$; $w$ inferred as conductance | Gate-current correlation; layer and retention controls [19] |
| Ion population in crystalline versus amorphous regions | Mixed-morphology organic OECT | Selective doping and trapping set volatile or nonvolatile mode | $I_D$ under standardized read; retained $G$ | Structure-dependent mode switching [20] |
| Ion profile; electrochemical doping fraction | Ion-gel/P3HT vertical synapse | Ions penetrate P3HT and relax or remain after pulses | Postsynaptic $I_D$; sampled current or $G$ | Electrical and geometry evidence; profile inferred [21] |

## S6. Electrical projections under voltage sweeps

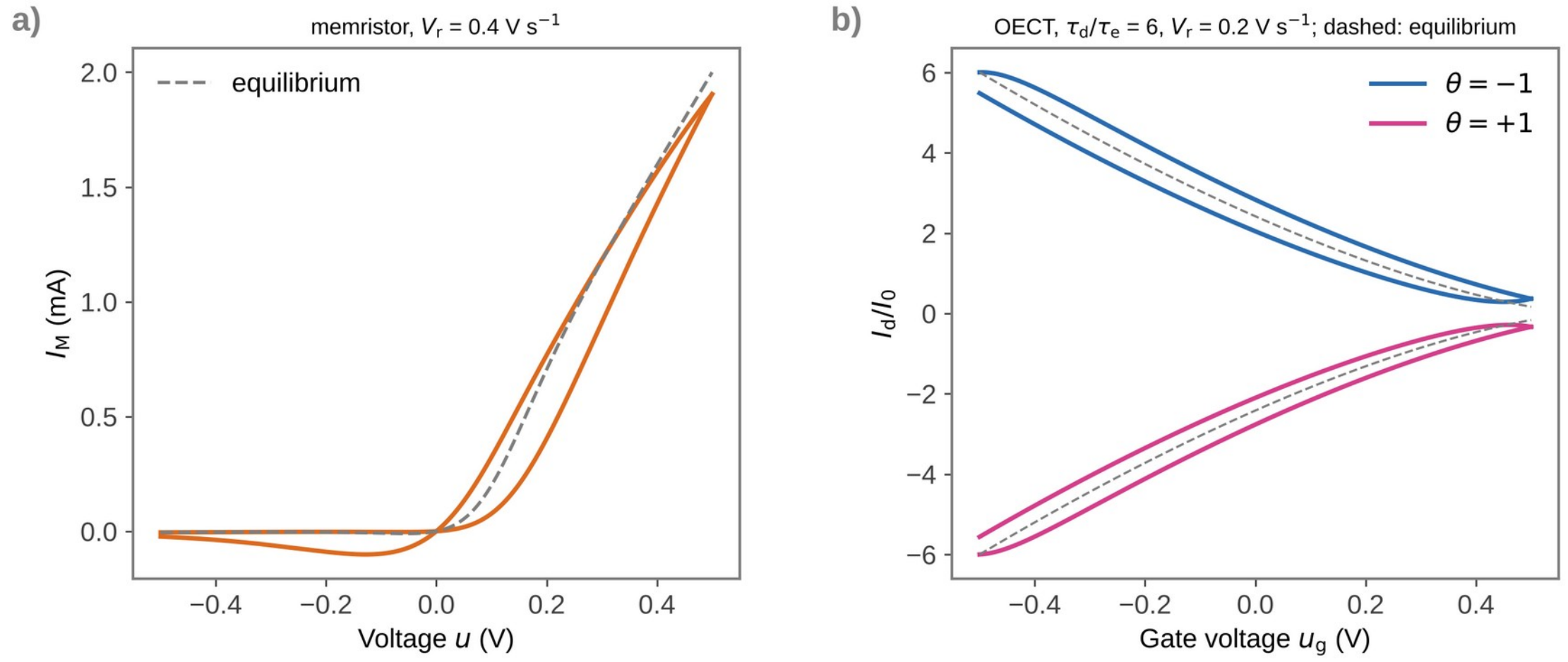

**Figure S3.** Voltage-sweep projections of the memristor and the accumulation-mode OECT. (a) Memristor current loop computed with Eq. (6) of the main text and Eq. (S1) at a sweep rate of 0.4 V s$^{-1}$; the dashed line is the equilibrium response, and the explicit voltage factor forces the current to vanish at zero bias. (b) OECT transfer curves, $I_d/I_0$, for $\theta=-1$ and $\theta=+1$ at $\tau_d/\tau_e=6$ and 0.2 V $s^{-1}$; dashed lines are the equilibrium responses. Both traces arise from the same charging equation, Eq. (8) of the main text; $\theta$ changes the signed projection of the state into the drain current, not the charging equation.

Fig. S3 extends the pulse comparison of the main text to voltage sweeps. In the memristor, the delayed coordinate modifies the conductance while the terminal voltage both drives and probes the device. In the OECT, the gate sweep drives the channel away from equilibrium and the drain circuit observes the resulting charging trajectory. The loop geometry consequently reports the selected state-to-current projection rather than the internal coordinate alone.

The two OECT traces in Fig. S3(b) are not different charging mechanisms. They are two signed electrical projections of the same delayed state $A(t)$. The $\theta=-1$ convention gives the positive pulse-current envelope used in Fig. S1, whereas $\theta=+1$ reverses the signed contribution of channel charging and permits the null-readout condition discussed in Section 4.1 of the main text and Fig. S2.

## S7. Accumulation-mode OECT: pulse direction, polarity factor, and time-scale ratio

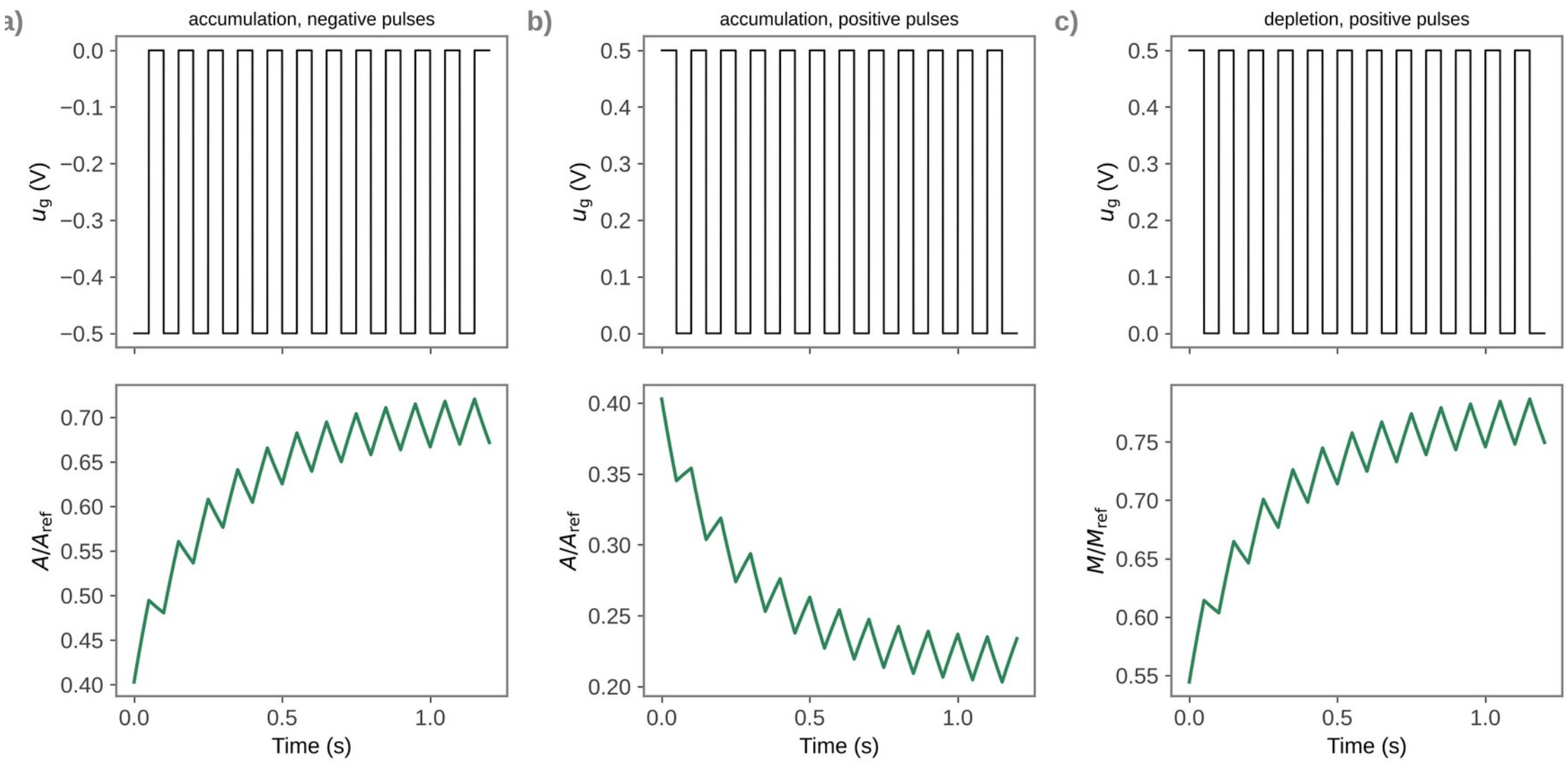

**Figure S4.** Gate-pulse protocols and state trajectories used in the supplementary series. (a) Negative gate pulses and the accumulation-mode charging coordinate $A/A_{ref}$; (b) positive gate pulses and $A/A_{ref}$; (c) positive gate pulses and the depletion coordinate $M/M_{ref}$, computed with Eqs. (S2) and (S3). The same dimensionless coordinates are used in the current calculations.

### S7.1. Complete accumulation-mode pulse-response series

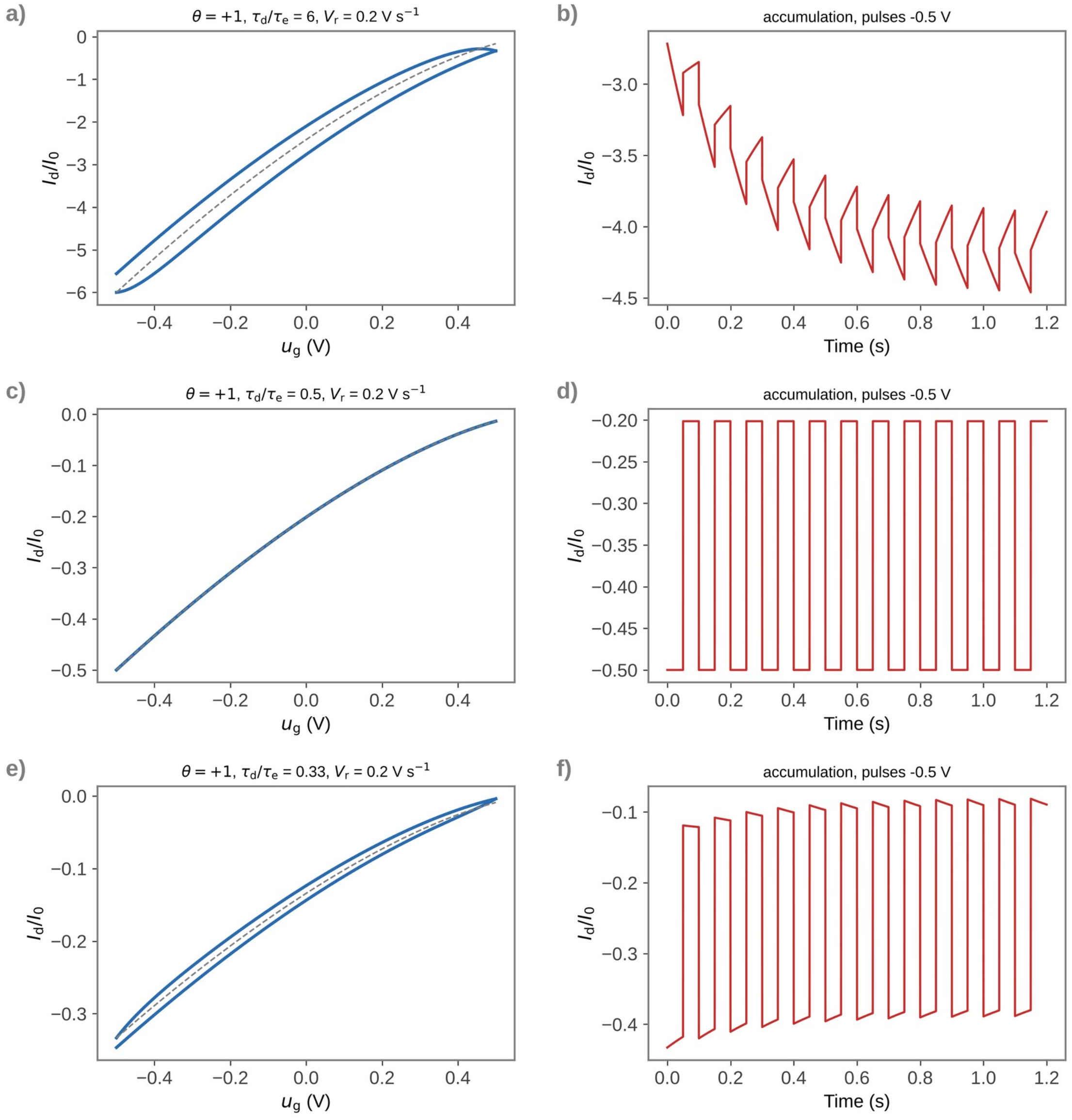

**Figure S5.** Accumulation-mode OECT under negative gate pulses for $\theta=+1$. Rows (a,b), (c,d), and (e,f) correspond to $\tau_d/\tau_e=6$, 0.5, and 0.33 ($\tau_e=50$, 600, and 900 ms at $\tau_d=300$ ms). Panels (a,c,e) show transfer curves and panels (b,d,f) the drain-current pulse responses, in units of $I_0$. The row at 0.5 satisfies the compensation condition of Eq. (15) of the main text and loses the pulse-to-pulse expression of the delayed state.

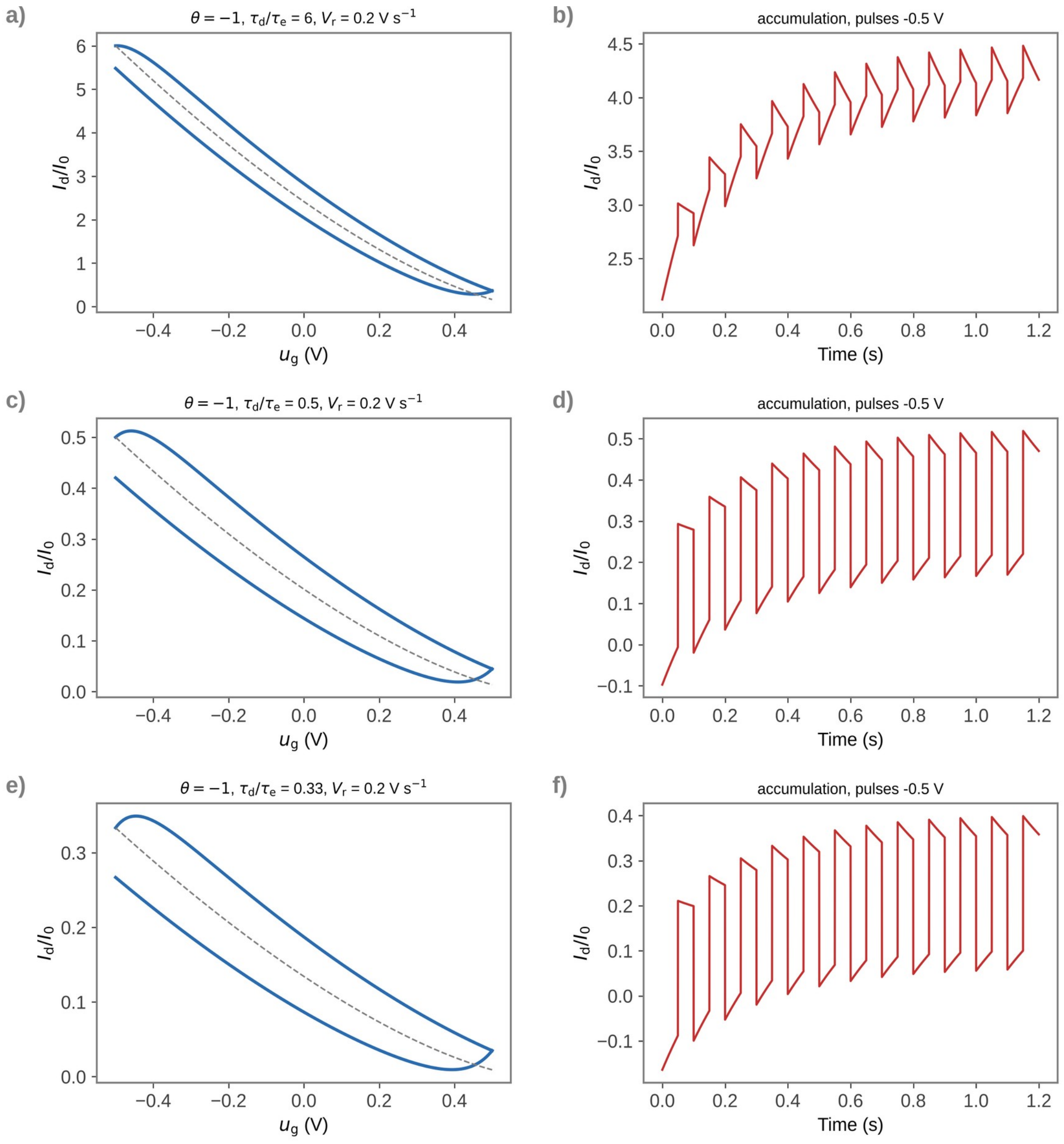

**Figure S6.** Accumulation-mode OECT under negative gate pulses for $\theta = -1$, with the same layout and time-scale ratios as Fig. S5: transfer curves (a,c,e) and pulse responses (b,d,f). The current retains a finite delayed-state contribution for all positive time-scale ratios.

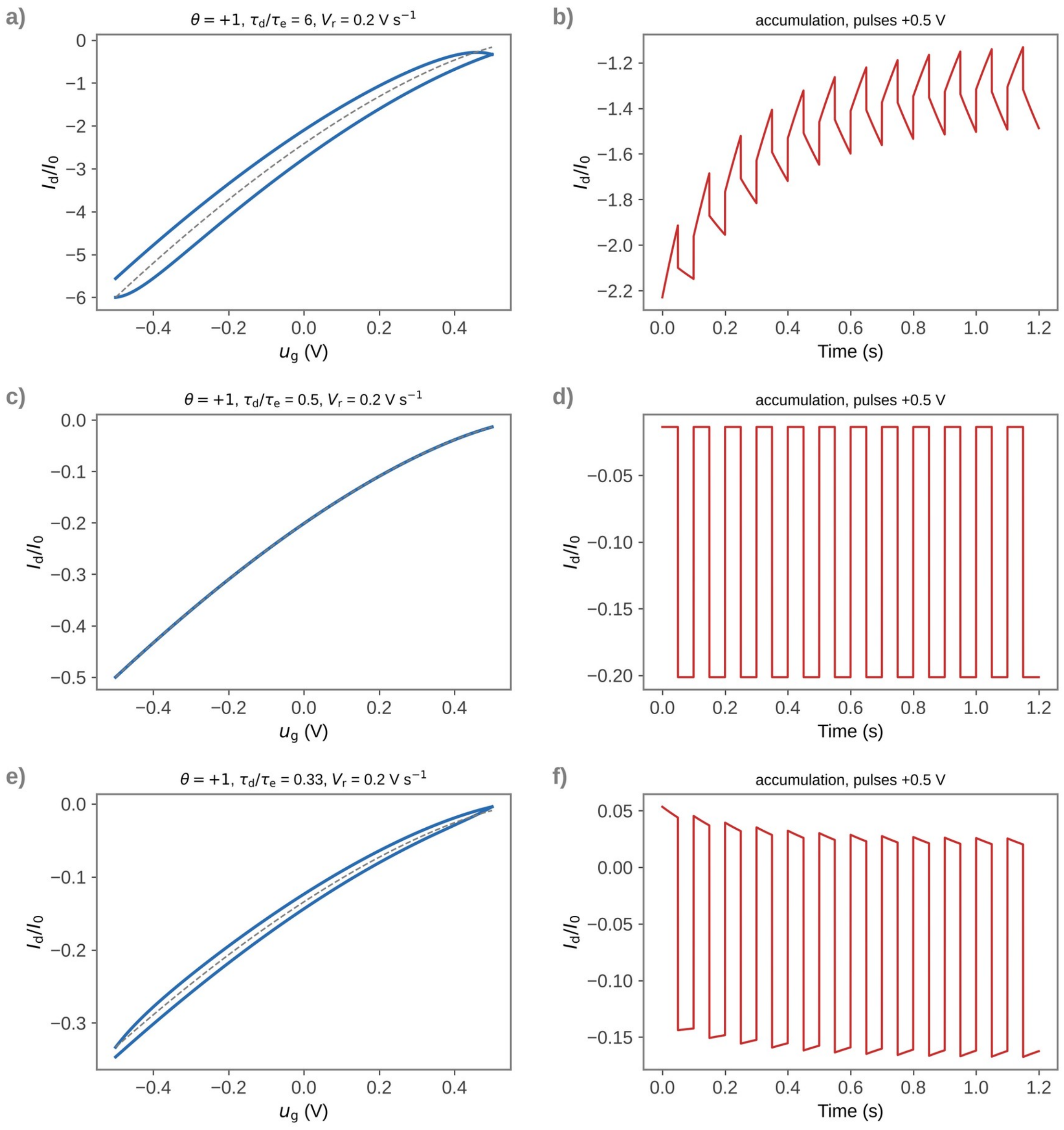

**Figure S7.** Accumulation-mode OECT under positive gate pulses for $\theta = +1$: transfer curves (a,c,e) and drain-current pulse responses (b,d,f) for $\tau_d/\tau_e = 6$, 0.5, and 0.33. The row at 0.5 provides the complementary pulse-direction example of the compensation condition.

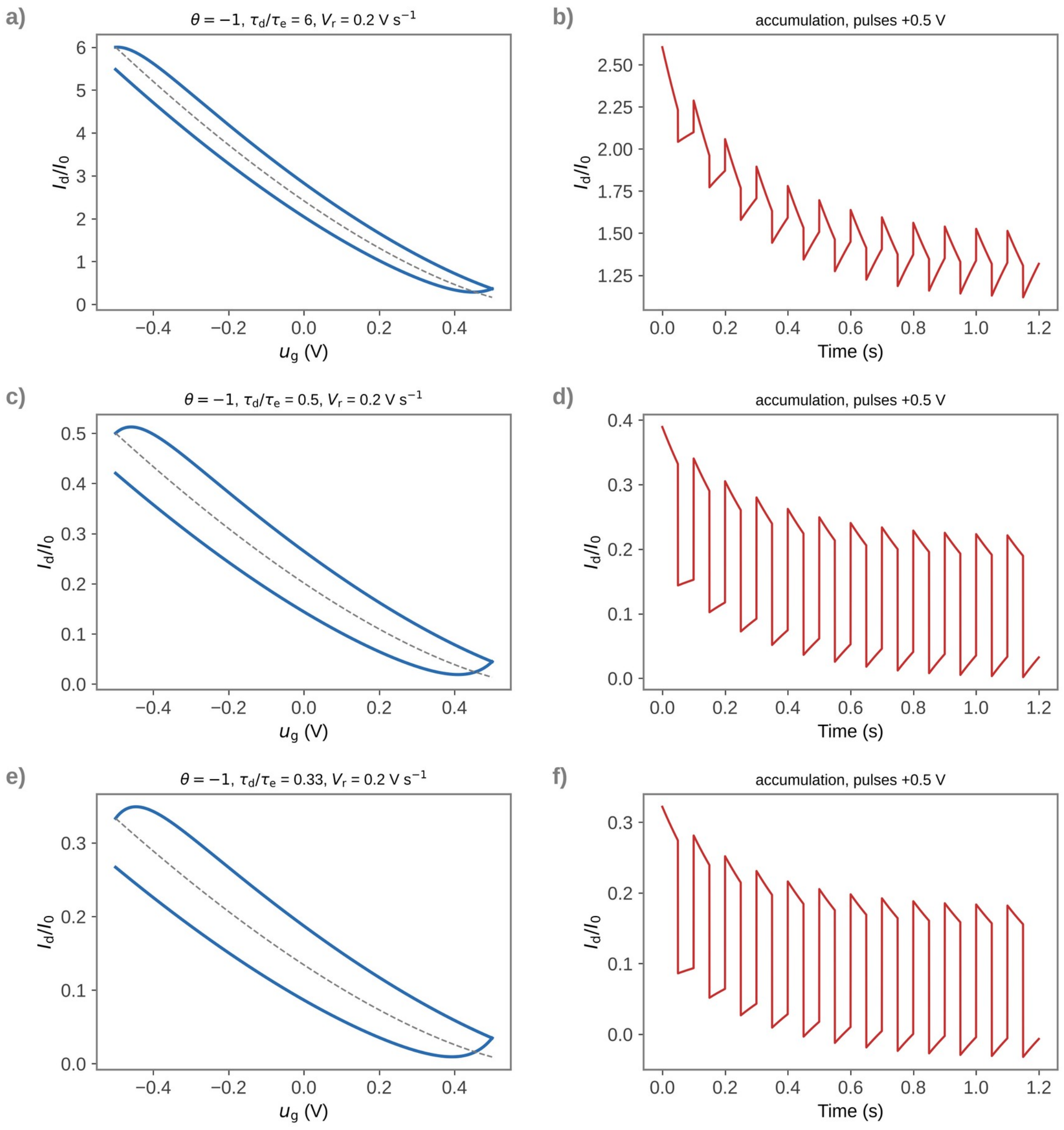

**Figure S8.** Accumulation-mode OECT under positive gate pulses for $\theta = -1$: transfer curves (a,c,e) and drain-current pulse responses (b,d,f) for $\tau_d/\tau_e = 6$, 0.5, and 0.33. Comparison with Fig. S7 separates the charging trajectory selected by the gate protocol from its signed projection into the drain current.

## S8. Interpretation of the extended null-readout series

Figures S5–S8 provide two controls for the state-observable interpretation of Section 4.1 of the main text. Changing the gate-pulse direction changes the equilibrium trajectory selected for $A$, whereas changing $\theta$ changes how that trajectory is projected into the measured drain current. These operations act on different parts of the state-readout relation.

For $\theta=+1$, $f=0.50$, and $r=0.50$, the coefficient multiplying the delayed state in Eq. (14) of the main text vanishes. The corresponding rows in Figs. S5 and S7 show that the selected drain-current observable loses its pulse-to-pulse delayed-state contribution, while Fig. S4 confirms that the charging coordinate continues to evolve under the gate protocol. For $\theta=-1$, the two state-dependent contributions have the same sign over the physical range $f>0$ and $r>0$, so the cancellation is absent in Figs. S6 and S8.

## S9. Depletion-mode extension: derivation and simulations

### S9.1. Depletion-mode state and current equations

The depletion-mode derivation tests whether the distinction between an internal charging coordinate and its electrical projection depends on the initially undoped accumulation-mode reference. The same construction applies when the channel is initially conducting, provided that the charging coordinate is defined relative to the initial carrier population.

A depletion-mode OECT differs from the accumulation-mode case because the channel is already electronically conducting at the reference gate potential. Let $P_0$ denote the mobile electronic carrier population in the initially doped channel. For a p-type depletion-mode device, $P_0$ corresponds to the equilibrium hole population before gate-induced de-doping. Gate-driven ionic charging reduces the mobile carrier population. We describe this process with the depletion coordinate $M$, defined by

$$P=P_0-M \qquad \text{(S8)}$$

The variable $M$ is a gate-induced depletion coordinate. It represents the ionic-electronic charging process that neutralizes or removes mobile carriers from the initially conducting channel. It is not an additional memory variable. The state dynamics are governed by relaxation toward a gate-voltage-dependent equilibrium,

$$\tau_d \frac{dM}{dt}=M_{eq}(u_g)-M \qquad \text{(S9)}$$

where $u_g$ is the gate voltage, $M_{eq}(u_g)$ is the equilibrium depletion state selected by the gate, and τd is the ionic charging time. The same ionic relaxation time controls the evolution of the depletion-mode state as in accumulation mode. What changes is the relation between the charging state and the mobile electronic population: increasing $M$ decreases $P$.

The drain current is obtained by combining the electronic readout of the remaining mobile carrier population with the dynamic contribution associated with relaxation of the depletion state,

$$I_d^{dep} = -\theta \frac{qL}{\tau_e}\left(P_0 - M\right) + qLf\frac{dM}{dt} \quad \text{(S10)}$$

Here, $\tau_e$ is the electronic transport time, $\theta$ determines the signed projection of the electronic channel state into the drain current, and f is the dimensionless partition factor of the relaxation contribution. The sign of the dynamic term differs from that in accumulation mode because the depletion coordinate increases while the mobile carrier population decreases.

Using Eq. (S9), Eq. (S10) can be written as

$$I_d^{dep} = -\theta \frac{qL}{\tau_e}\left(P_0 - M\right) + \frac{qLf}{\tau_d}\left[M_{eq}\left(u_g\right) - M\right] \quad \text{(S11)}$$

or, equivalently,

$$I_d^{dep} = -\theta \frac{qL}{\tau_e} P_0 + \frac{qL}{\tau_d}\left[\left(\theta \frac{\tau_d}{\tau_e} - f\right) M + f M_{eq}\left(u_g\right)\right] \quad \text{(S12)}$$

Eq. (S12) separates a constant current offset associated with the initially conducting channel from the gate-dependent current response. The first term does not affect the transient or hysteretic response. The remaining terms contain a contribution proportional to the delayed internal state $M$ and a contribution proportional to the gate-defined equilibrium state $M_{eq}$.

At a fixed gate voltage, the depletion state relaxes toward $M_{eq}$, and the steady-state drain current becomes

$$I_{d,ss}^{dep} = -\theta \frac{qL}{\tau_e}\left[P_0 - M_{eq}\left(u_g\right)\right] \quad \text{(S13)}$$

Thus, the steady-state current is determined by the remaining mobile carrier population. A larger depletion state $M_{eq}$ corresponds to a smaller conducting population $P$.

For a gate pulse returning to a baseline voltage $u_{g,b}$, define

$$M_b = M_{eq}\left(u_{g,b}\right) \quad \text{(S14)}$$

After the pulse is removed, the depletion state relaxes as

$$M(t) = M_b + \left[M(t_0) - M_b\right] e^{-(t-t_0)/\tau_d} \quad \text{(S15)}$$

where $t_0$ is the instant at which the gate returns to baseline. The drain current approaches the finite baseline value

$$I_{d,b}^{dep} = -\theta \frac{qL}{\tau_e}\left(P_0 - M_b\right) \quad \text{(S16)}$$

The finite current between gate pulses is maintained by the drain-source bias and is not,

by itself, evidence of nonvolatile retention. In the reversible-charging model, the pulse-induced deviation from the baseline depletion state relaxes with the ionic time $\tau_d$. Equation (S12) also shows that the coefficient of $M$ vanishes at the same condition $\tau_d/\tau_e = f/\theta$ as in the accumulation-mode model. The internal depletion coordinate can therefore remain dynamic while the selected drain-current observable becomes locally insensitive to it.

### S9.2. Depletion-mode polarity and time-scale series

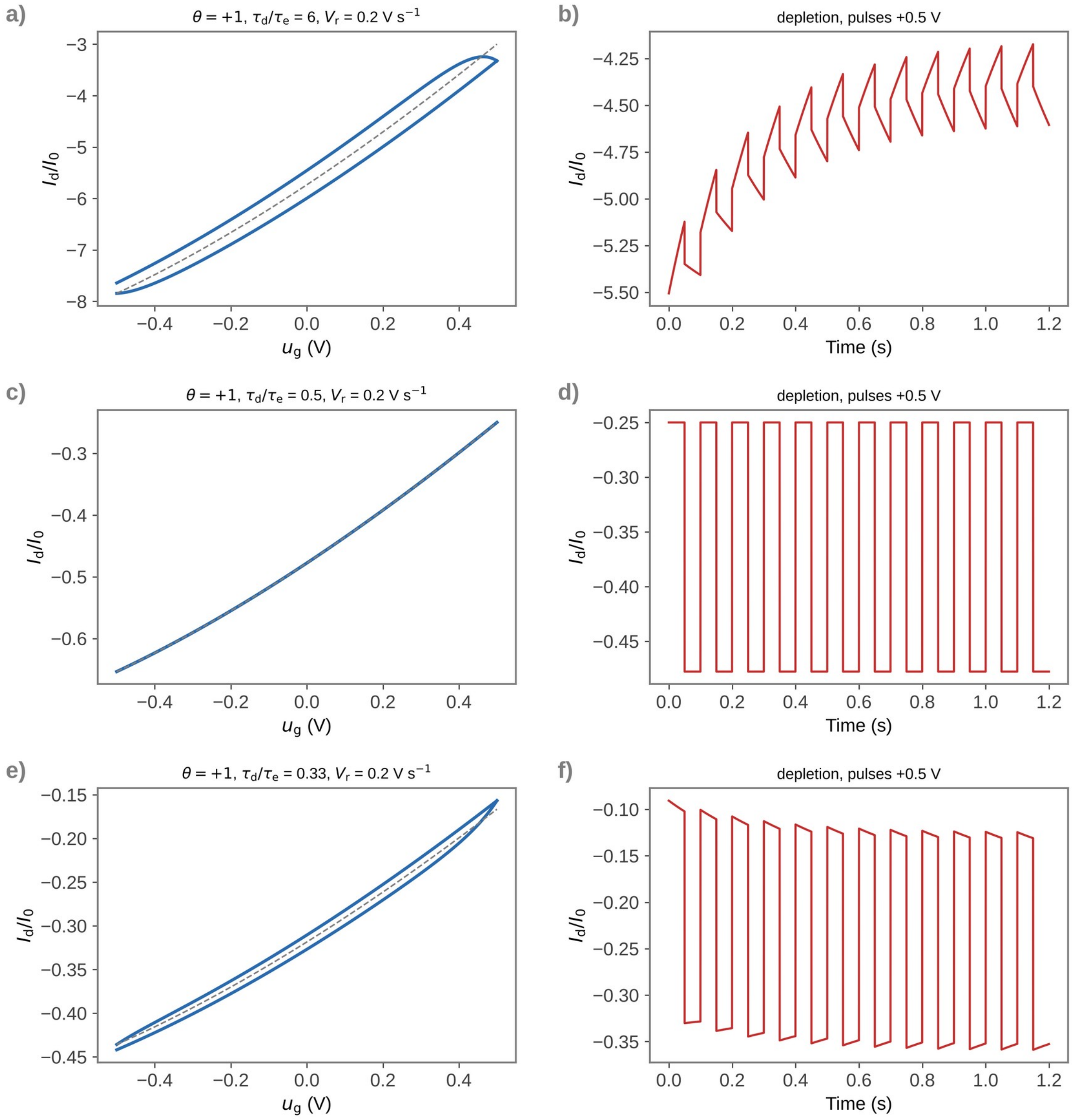

**Figure S9.** Depletion-mode OECT under positive gate pulses for $\theta = +1$: transfer curves (a,c,e) and pulse responses (b,d,f) for $\tau_d/\tau_e = 6$, 0.5, and 0.33, with currents in units of $qLM_{ref}/\tau_d$. The row at 0.5 illustrates the depletion-mode counterpart of the compensation derived from Eq. (S12).

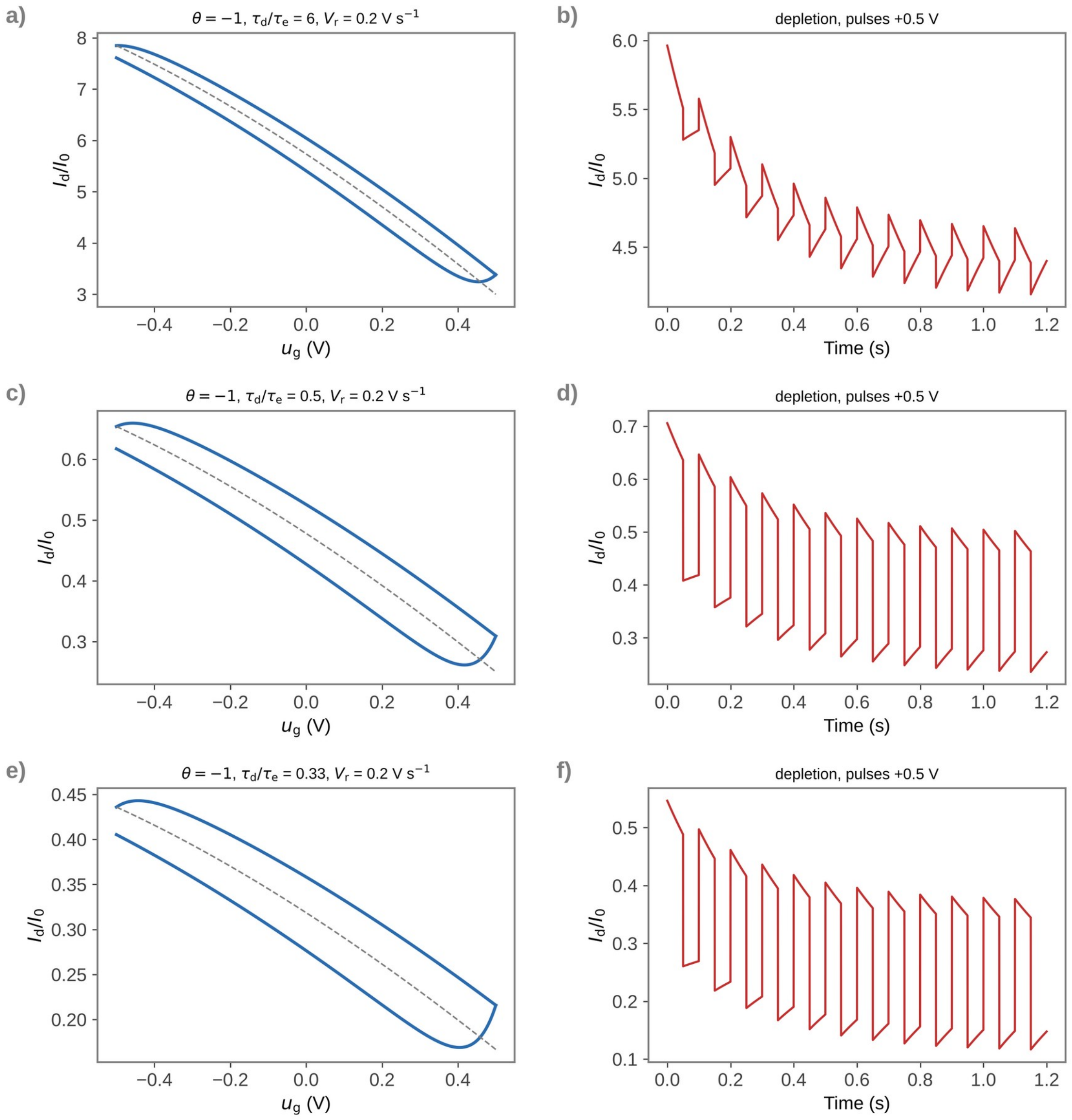

**Figure S10.** Depletion-mode OECT under positive gate pulses for $\theta = -1$, with the same layout and time-scale ratios as Fig. S9. The same depletion dynamics are projected into the opposite signed drain current, without a compensation point.

Together, Figs. S9 and S10 show that the state-observable distinction developed in the main text is independent of whether gate charging increases or decreases the mobile carrier population. Accumulation and depletion modes differ in their reference carrier population and current sign structure, while both remain governed by a delayed charging coordinate and a separate electrical observation map.

## S10. Robustness of the compensation

The exact cancellation of Eq. (14) of the main text holds for the affine single-coordinate model. Fig. S11 evaluates, for the two gate histories of Fig. 3 of the main text read at the nominal compensation condition $\tau_d/\tau_e = f = 0.5$, the residual drain-current contrast produced by two departures from that model. (a) A second charging mode with time constant $\tau_{d,2} = 1.5$ s carries a fraction 0.1 or 0.2 of the equilibrium charge, and the charging current of both modes is partitioned with the same factor $f$; with a sensitivity of terminal $k$ to mode $j$ given by Eq. (S17), the slow mode is not compensated at the read condition of the fast mode. (b) The partition factor depends on the charging state, $f(A) = f_0\left[1+\epsilon\left(A/A_{ref} - 0.5\right)\right]$ with $f_0 = 0.5$ and $\epsilon = 0.2$ or 0.5. In all cases the residual contrast stays below the resolution $\delta I_{res} = 0.01 I_0$ used in Fig. 3 of the main text, approaching it only for the largest slow-mode fraction. The affine case (fraction 0, $\epsilon = 0$) gives zero contrast and is not visible on the logarithmic scale. Hiding is therefore a statement about a residual contrast relative to a resolution, which remains valid for moderate departures from the affine model and fails when a slowly charging population carries a large share of the stored charge.

$$C_{k,j} = -qL\left(\frac{\theta}{\tau_e} - \frac{s_k f_k}{\tau_{d,j}}\right) \tag{S17}$$

is the sensitivity of terminal $k$ to a charging mode $A_j$ with ionic time $\tau_{d,j}$; a single read condition cannot cancel all modes.

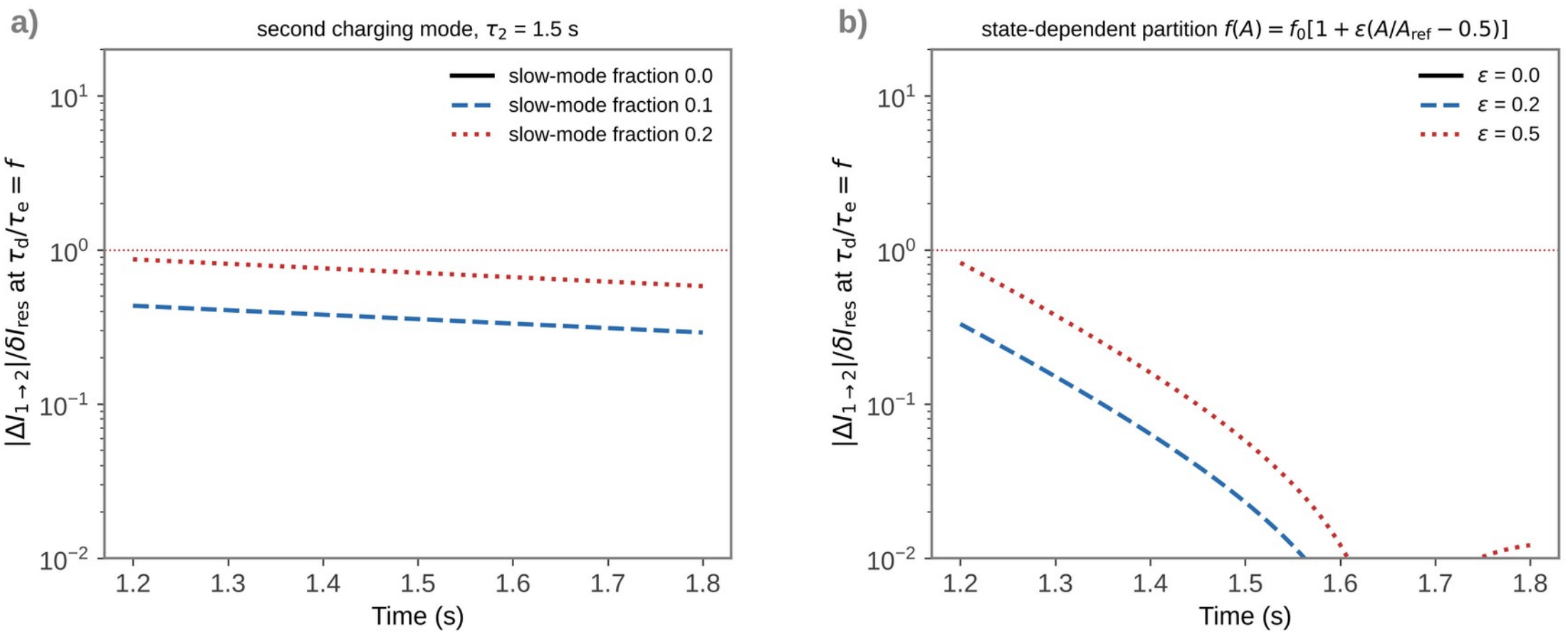

**Figure S11.** Residual drain-current contrast between the two histories of Fig. 3 of the main text, divided by $\delta I_{res} = 0.01 I_0$, at the nominal compensation condition $\tau_d/\tau_e = f$ during the read ($t \geq 1.2$ s). (a) Second charging mode with $\tau_{d,2} = 1.5$ s carrying fractions 0.1 (dashed) and 0.2 (dotted) of the charge. (b) State-dependent partition factor with $\epsilon = 0.2$ (dashed) and 0.5 (dotted). The dotted horizontal line marks the resolution.

## S11. Code availability

The Python scripts used for all model figures (make_figs4.py for Fig. 3 and Figs. S1

and S2, make_figs6.py for Figs. 1 and 2, and make_si_figs.py for Figs. S3–S11) are provided with the article. The experimental panels of Fig. 2 are reproduced from the cited works.